\documentclass{aa}
\include{amssymb}
\usepackage{graphicx}
\usepackage{txfonts}
\usepackage{natbib}
\usepackage{hyperref}
\hypersetup{colorlinks=true, citecolor=blue}
\usepackage{graphicx}
\usepackage{txfonts}
\usepackage{rotating}
\bibpunct{(}{)}{;}{a}{}{,}

\begin{document} 
\title{Measuring star formation with resolved observations: the test case of M33}
\author{M. Boquien\inst{1,2} \and D. Calzetti\inst{3} \and S. Aalto\inst{4} \and A. Boselli\inst{2} \and J. Braine\inst{5} \and V. Buat\inst{2} \and F. Combes\inst{6} \and F. Israel\inst{7} \and C. Kramer\inst{8} \and S. Lord\inst{9} \and M. Rela\~no\inst{10} \and E. Rosolowsky\inst{11} \and G. Stacey\inst{12} \and F. Tabatabaei\inst{13} \and F. van der Tak\inst{14,15} \and P. van der Werf\inst{7} \and S. Verley\inst{10} \and M. Xilouris\inst{16}}
\authorrunning{M. Boquien et al.}
\institute{
  Institute of Astronomy, University of Cambridge, Madingley Road, Cambridge CB3 0HA, UK \email{mboquien@ast.cam.ac.uk}
  \and Aix Marseille Universit\'e, CNRS, LAM (Laboratoire d'Astrophysique de Marseille) UMR 7326, 13388, Marseille, France
  \and Department of Astronomy, University of Massachusetts, Amherst, MA 01003, USA
  \and Department of Earth and Space Sciences, Chalmers University of Technology, Onsala Observatory, 43994, Onsala, Sweden
  \and Laboratoire d'Astrophysique de Bordeaux, Universit\'e de Bordeaux and CNRS UMR 5804, 33271, Floirac, France
  \and Observatoire de Paris, LERMA, CNRS, 61 Av. de l'Observatoire, 75014, Paris, France
  \and Sterrewacht Leiden, Leiden University, PO Box 9513, 2300 RA, Leiden, The Netherlands
  \and Instituto Radioastronomia Milimetrica, 18012, Granada, Spain
  \and Infrared Processing and Analysis Center, California Institute of Technology, MS 100--22, Pasadena, CA 91125, USA
  \and Department F\'isica Te\'orica y del Cosmos, Universidad de Granada, 18071, Granada, Spain
  \and Department of Physics, University of Alberta, 2--115 Centennial Centre for Interdisciplinary Science, Edmonton AB, Canada
  \and Department of Astronomy, Cornell University, Ithaca, NY 14853, USA
  \and Max-Planck-Institut f\"ur Astronomie, K\"onigstuhl 17, 69117, Heidelberg, Germany
  \and SRON Netherlands Institute for Space Research, Landleven 12, 9747 AD, Groningen, The Netherlands
  \and Kapteyn Astronomical Institute, University of Groningen, The Netherlands
  \and Institute for Astronomy, Astrophysics, Space Applications and Remote Sensing, National Observatory of Athens, 15236, Athens, Greece
}
\date{}

\abstract{Measuring star formation at a local scale is important to constrain star formation laws. Yet, it is not clear whether and how the measure of star formation is affected by the spatial scale at which a galaxy is observed.}{We want to understand the impact of the resolution on the determination of the spatially resolved star formation rate (SFR) and other directly associated physical parameters such as the attenuation.}{We have carried out a multi--scale, pixel--by--pixel study of the nearby galaxy M33. Assembling FUV, H$\alpha$, 8~$\mu$m, 24~$\mu$m, 70~$\mu$m, and 100~$\mu$m maps, we have systematically compared the emission in individual bands with various SFR estimators from a resolution of 33~pc to 2084~pc.}{We have found that there are strong, scale--dependent, discrepancies up to a factor 3 between monochromatic SFR estimators and H$\alpha$+24~$\mu$m. The scaling factors between individual IR bands and the SFR show a strong dependence on the spatial scale and on the intensity of star formation. Finally, strong variations of the differential reddening between the nebular emission and the stellar continuum are seen, depending on the specific SFR (sSFR) and on the resolution. At the finest spatial scales, there is little differential reddening at high sSFR. The differential reddening increases with decreasing sSFR. At the coarsest spatial scales the differential reddening is compatible with the canonical value found for starburst galaxies.}{Our results confirm that monochromatic estimators of the SFR are unreliable at scales smaller than 1~kpc. Furthermore, the extension of local calibrations to high redshift galaxies presents non--trivial challenges as the properties of these systems may be poorly known.}


\maketitle

\section{Introduction\label{sec:intro}}
As we observe galaxies across the Universe, their evolution from highly disturbed proto--galaxies at high redshift to the highly organised systems common in the zoo of objects we see in the nearby Universe is striking. One of the most important processes that drives this evolution is the transformation of the primordial gas reservoir into stars, which form heavy elements that are ejected into the intergalactic medium during intense episodes of feedback. In other words, if we want to understand galaxy formation and evolution across cosmic times, we need to understand the process of star formation in galaxies. To do so, it is paramount to be able to measure star formation as accurately as possible.

The most direct way to trace star formation is through the photospheric emission of massive stars with lifetimes of up to $\sim100$~Myr, which dominate the ultraviolet (UV) energy budget of star--forming galaxies. An indirect star formation tracer is the H$\alpha$ recombination line (or any other hydrogen recombination line) from gas ionised by the most massive stars that are around for up to $\sim10$~Myr. However, both the UV emission and the H$\alpha$ line are severely affected by the presence of dust, which absorbs energetic photons and reemits their energy at longer wavelengths. From the inception of the far--infrared era with the launch of IRAS \cite[Infrared Astronomical Satellite,][]{neugebauer1984a}, the emission of the dust has been used as a powerful tracer of star formation from local galaxies up to high--redshift objects, resulting in a tremendous progress of our understanding of galaxy evolution in general and of the physical processes of star formation in particular.

The launch of the \textit{Herschel Space Observatory} \citep{pilbratt2010a} has opened new avenues for the investigation of star formation in the far--infrared not only in entire galaxies, but also within nearby galaxies at scales where physical processes such as heating and cooling are localised. \textit{Herschel} matches the angular resolution of 5--6\arcsec of \textit{Spitzer} in the mid--infrared \citep{fazio2004a,rieke2004a}, and of GALEX in the UV \citep[Galaxy Evolution Explorer,][]{martin2005a}.

Yet, measuring local star formation in galaxies remains an important challenge. For instance, \cite{kennicutt2007a,bigiel2008a} found seemingly incompatible star formation laws with the same dataset. Such a difference could be due to the distinct ways star formation is measured in galaxies \citep{liu2011a}, different SFR estimates leading to variations of 10--50\% of the molecular gas depletion timescale \citep{leroy2012a,leroy2013a}.

The measurement of star formation relies upon three main assumptions.
\begin{itemize}
 \item First of all, a well--defined and fully sampled initial mass function (IMF) is assumed. This is necessary to relate the measured power output from massive, short--lived stars to the total mass of the stellar population of the same age. Massive stars only account for a minor fraction of the total mass of stellar populations, even in the youngest star--forming regions, which contain the highest proportion of such stars.
 \item Star--formation--tracing bands need to be sensitive mainly to the most recent episode of star formation. Contamination from emission unrelated to recent star formation, such as active nuclei and older stellar populations, needs to be negligible.
 \item A well--defined star formation history. Too few star--forming regions would induce rapid variations of the SFR with time.
\end{itemize}
These assumptions, which are not exhaustive, may already be problematic for some entire galaxies \citep{boselli2009a}. At small scales, they are unlikely to hold true across an entire spiral disk.

If we want to understand star formation laws in the era of resolved observations, it is therefore crucial to understand when, how, and from which spatial scale we can measure star formation reliably. In particular, we need to understand how star formation tracers relate to each other in galaxies from the finest spatial scale, at which H\,\textsc{ii} regions are resolved, to large fractions of a spiral disk. Recent results show a systematic variation of star formation tracers with spatial scale, which could be due to the presence of diffuse emission unrelated to recent star formation \citep{li2013a}: $\sim20-30$\% of the FUV luminosity from a galaxy is due to stars older than 100~Myr \citep{johnson2013a,boquien2014a} and a 30\% to 50\% of H$\alpha$ is diffuse \citep{thilker2005b, crocker2013a}. Measuring local star formation is made even more difficult by the fact that indirect tracers of star formation (the ionised gas and dust emission) may not be spatially coincident with the direct tracer of star formation, the UV emission \citep[][]{calzetti2005a,relano2009a,verley2010b,louie2013a, relano2013a}. Such offsets can also be seen in the Milky Way in NGC~3603, Carina, or the OB associations in Orion for instance. This questions the real meaning of SFR measurements at local scales.

These offsets along with other processes such as stochastic sampling of the IMF or the insufficient number of star--forming regions and/or molecular clouds in a given region could be one of several reasons for the observational breakdown of the Schmidt--Kennicutt law on scales of the order of $\sim$100--300~pc \citep{calzetti2012a}, which has been found in M33 by \cite{onodera2010a} and \cite{schruba2010a}. The complex interplay between various processes at the origin of the breakdown of the Schmidt--Kennicutt law at small spatial scales has recently been analysed by \cite{kruijssen2014a}.

With the availability of resolved observations of high--redshift objects with ALMA and the JWST by the end of the decade, understanding whether and how we can measure star formation at local scales is also of increasing importance. We address this question through a detailed study of star formation tracers at all scales in the nearby late--type galaxy M33. Thanks to its proximity \citep[840~kpc, corresponding to 4.07 pc/\arcsec,][]{freedman1991a}, relatively low inclination \citep[56$^\circ$,][]{regan1994a}, and large angular size (over $1^\circ$ across), M33 is an outstanding galaxy for such a study. It has been a popular target for a large number of multi-wavelengths observations and surveys in star--formation tracing bands from the FUV with the GALEX Nearby Galaxies Survey \citep[NGS,][]{gildepaz2007a}, to the FIR with \textit{Herschel} in the context of the HerM33es survey \citep{kramer2010a}, including \textit{Spitzer} mid--infrared data \citep{verley2007a} as well as H$\alpha$ narrow--band imaging \citep{hoopes2000a}.

In Sect.~\ref{sec:obs} we present the data, including new observations recently obtained by our team, and how data processing was carried out. We compare various SFR estimators at different scales in Sect.~\ref{sec:comp-SFR-scales}. We examine in detail the properties of dust emission with scale in order to measure the SFR from monochromatic infrared bands in Sect.~\ref{sec:scale-dep-estimators}. We investigate the relative fraction of attenuated and unattenuated star formation with scale in Sect.~\ref{sec:att-frac}. Finally, we discuss our results in Sect.~\ref{sec:discussion} and we conclude in Sect.~\ref{sec:conclusion}.

\section{Observations and data processing\label{sec:obs}}

\subsection{Observations}

To carry out this study, we consider the main star formation tracers used in the literature: the emission from young, massive stars in the FUV, the ionised gas recombination line H$\alpha$, and the emission of the dust at 8~$\mu$m, 24~$\mu$m, 70~$\mu$m, and 100~$\mu$m. We do not explore dust emission beyond 100~$\mu$m as prior work  has shown that longer wavelengths are not good tracers of star formation \citep{bendo2010a,bendo2012a,boquien2011a} and the scale sampled with \textit{Herschel} becomes coarser. We also forego radio tracers as they are not as widely used.

The FUV GALEX data from NGS were obtained directly from the GALEX website through \textsc{galexview}\footnote{\url{http://galex.stsci.edu/GalexView/}}. The observation was carried out on 25 November 2003 for a total exposure time of 3334~s.

H$\alpha$+[N\textsc{ii}] observations were carried out in November 1995 on the Burrel Schmidt telescope at Kitt Peak National Observatory. They consisted in 20 exposures of 900~s, each covering a final area of $1.75\times1.75$~deg$^2$. This map has been continuum--subtracted by scaling an off--band image using foreground stars. The observations and the data processing are analysed in detail in \cite{hoopes2000a}.

The \textit{Spitzer} IRAC 8~$\mu$m image sensitive to the emission of Polycyclic Aromatic Hydrocarbons (PAH) and the MIPS 24~$\mu$m image sensitive to the emission of Very Small Grains (VSG) were obtained from the NASA Extragalactic Database and have been analysed by \cite{hinz2004a} and \cite{verley2007a}.

The PACS data at 70~$\mu$m and 100~$\mu$m, which are sensitive to the warm dust heated by massive stars, come from two different programmes. The 100~$\mu$m image was obtained in the context of the \textit{Herschel} HerM33es open time key project \citep[][observation ID 1342189079 and 1342189080]{kramer2010a}. The observation was carried out in parallel mode on 7 January 2010 for a duration of 6.3~h. It consisted in 2 orthogonal scans at a speed of 20\arcsec/s, with a leg length of 70\arcmin. The 70~$\mu$m image was obtained as a follow--up open time cycle 2 programme (\textsc{OT2\_mboquien\_4}, observation ID 1342247408 and 1342247409). M33 was scanned on 25 June 2012 at a speed of 20\arcsec/s in 2 orthogonal directions over $50\arcmin$ with 5 repetitions of this scheme in order to match the depth of the 100~$\mu$m image. The total duration of the observation was 9.9~h. Reduced maps are available on the \textit{Herschel} user provided data product website\footnote{\url{http://www.cosmos.esa.int/web/herschel/user-provided-data-products}\label{ftnt:data}}.

\subsection{Additional data processing}

The GALEX data we obtained from \textsc{galexview} were already fully processed and calibrated, we therefore did not carry out any additional processing.

We corrected the continuum--subtracted H$\alpha$ map for [N\,\textsc{ii}] contamination, which according to \cite{hoopes2000a} accounts for 5\% of the H$\alpha$ flux in the narrow--band filter. We have also removed  subtraction artefacts caused by bright foreground stars. To do so we have used \textsc{iraf}'s \textsc{imedit} procedure, replacing these artefacts with data similar to that of the neighbouring background.

The \textit{Spitzer} IRAC and MIPS data we used were processed in the context of the Local Volume Legacy survey \citep[LVL,][]{dale2009a}. No further processing was performed.

\begin{figure*}[!htbp]
\includegraphics[width=\textwidth]{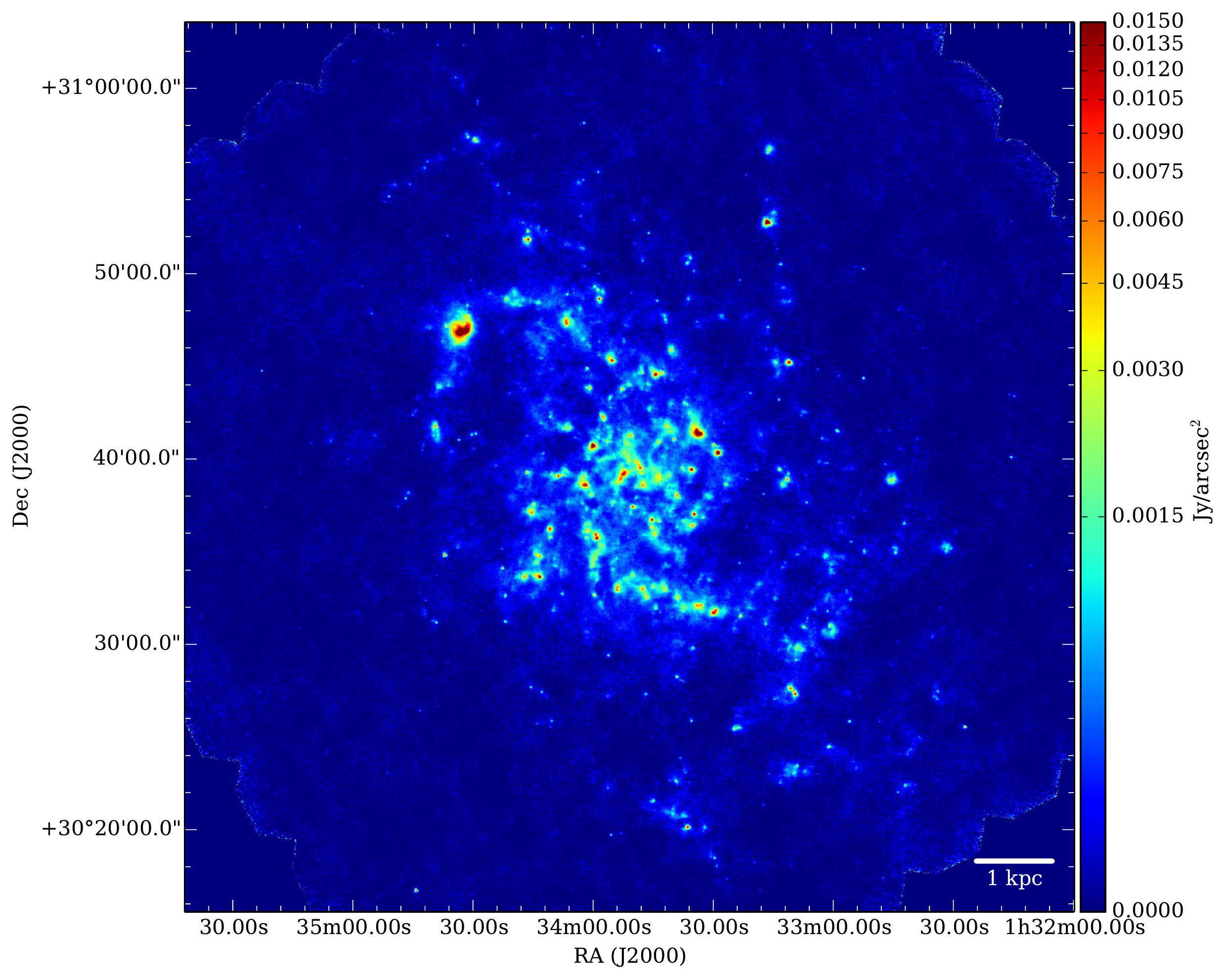}
\caption{Map of M33 at 70~$\mu$m obtained with the \textit{Herschel} PACS instrument in the context of a cycle 2 programme (\textsc{OT2\_mboquien\_4}, observation ID 1342247408 and 1342247409; the original map is available from the link given in footnote \ref{ftnt:data}). The image is in Jy/arcsec$^2$ and the colours follow an arcsinh scale indicated by the bar on the right side of the figure. The physical scale is indicated by the white line in the bottom--right corner of the figure, representing 1~kpc. Each pixel has a size of 1.4\arcsec.\label{fig:pacs70}}
\end{figure*}

Even though in the context of the HerM33es project we already reduced and published 100~$\mu$m data \citep{boquien2010b,boquien2011a}, these observations were processed with older versions of the data reduction pipeline. To work on a fully consistent set of \textit{Herschel} PACS data and to take advantage of the recent improvements of the pipeline, we have reprocessed the 100~$\mu$m from the HerM33es survey along with the new 70~$\mu$m data. To do so we have taken the raw data to level 1 with HIPE version 9 \citep{ott2010a}, flagging bad pixels, masking saturated pixels, adding pointing information, and calibrating each frame. In a second step, to remove the intrinsic $1/f$ noise of the bolometers and make the maps, we used the Scanamorphos software \citep{roussel2013a}, version 19. We present the new 70~$\mu$m map obtained for this project in Fig.~\ref{fig:pacs70}.

\subsection{Correction for the Galactic foreground extinction}

To correct the FUV and H$\alpha$ fluxes for the Galactic foreground extinction, we have used the \cite{cardelli1989a} extinction curve, including the \cite{odonnell1994a} update. We have assumed $\mathrm{E(B-V)=0.0413}$ as indicated by NASA/IPAC Infrared Science Archive's dust extinction tool from the \cite{schlegel1998a} extinction maps. This yields a correction of 0.34~mag in FUV and 0.11~mag in H$\alpha$.

\subsection{Astrometry}

To carry out a pixel--by--pixel analysis, it is important that the relative astrometric accuracy of all the bands is significantly better than the pixel size. A first visual inspection reveals a clear offset between the new 70~$\mu$m data we present in this paper and the 100~$\mu$m data presented in \cite{boquien2010b,boquien2011a}. When comparing the 70~$\mu$m and 100~$\mu$m images with the H$\alpha$ image of \cite{hoopes2000a}, we found that the 70~$\mu$m map corresponded more closely to the H$\alpha$ emission across the galaxy and was consistent with data at other wavelengths. We have therefore decided to shift the 100~$\mu$m image to match the 70~$\mu$m map astrometry. To determine the offset, we have compared the relative astrometry of the 160~$\mu$m images obtained in the context of HerM33es and the \textsc{OT2\_mboquien\_4} programme. As the 160~$\mu$m is observed in parallel with the 70~$\mu$m or the 100~$\mu$m, they have the same astrometry. The offset between the 160~$\mu$m maps between these two programmes is therefore the same as the offset between the 70~$\mu$m and the 100~$\mu$m maps. We have determined an offset of $\sim5\arcsec$ (4.83\arcsec\ towards the East and 1.25\arcsec\ towards the North) and applied this  to the 100~$\mu$m image. When comparing the corrected 100~$\mu$m band with the 8~$\mu$m and 24~$\mu$m images we can see small region--dependent offsets of the order of 1--2\arcsec. The variation of this offset from one region to another leads us to think that at least part of it reflects physical variations in the emission of the various dust components in M33. In addition, as we will see later, we carry out this study at a minimum pixel size 8\arcsec, much larger than any possible systematic offset. We conclude that the relative astrometry of our images is sufficient to reach our goals.

\subsection{Pixel--by--pixel matching}

\begin{figure}[!htbp]
\includegraphics[width=\columnwidth]{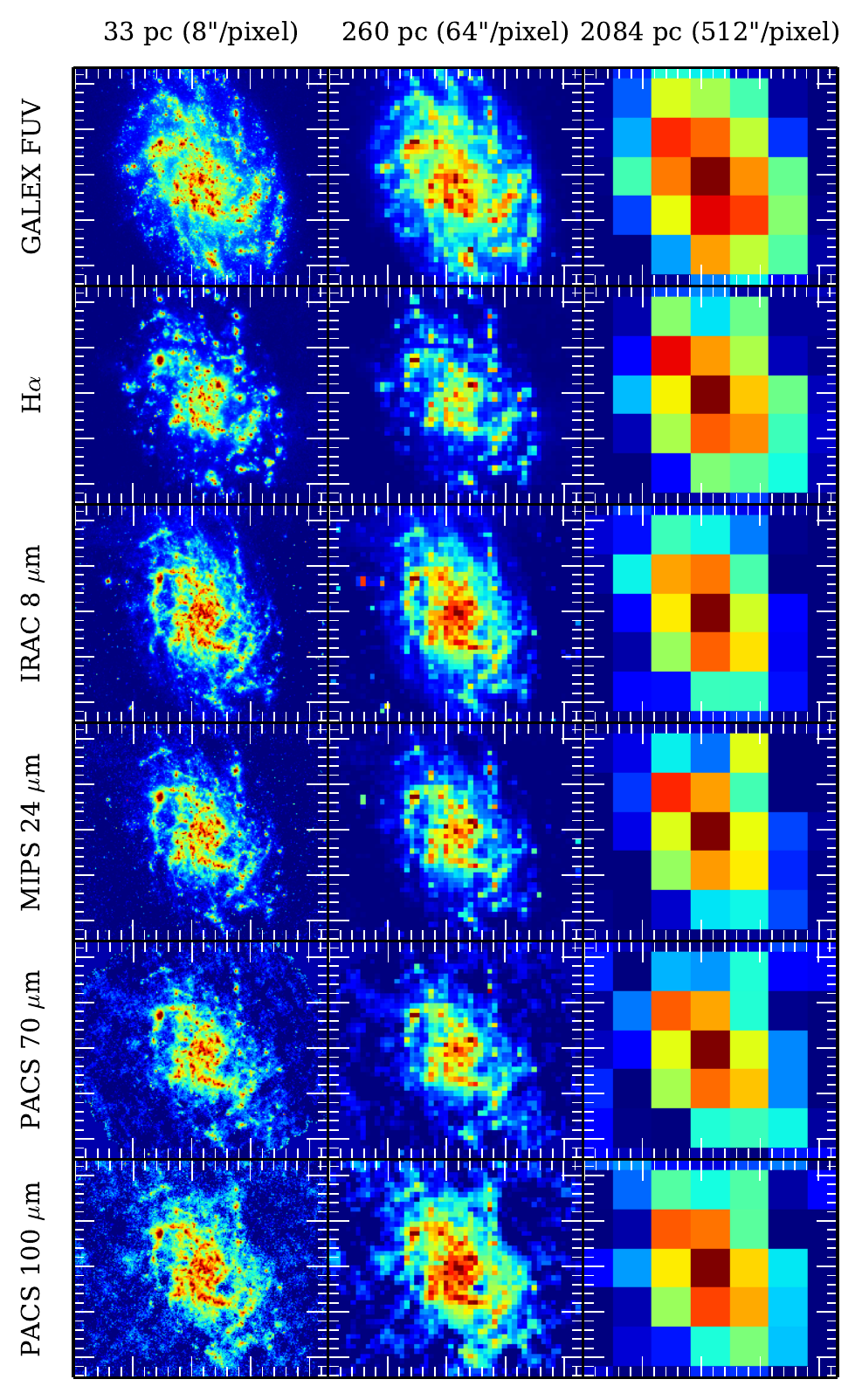}
\caption{Convolved images registered to a common reference frame at 33~pc (8\arcsec/pixel, left), 260~pc (64\arcsec/pixel, centre), and 2084~pc (512\arcsec/pixel, right). Each row represents a different star formation tracing band, from top to bottom: GALEX FUV, H$\alpha$, IRAC 8~$\mu$m, MIPS 24~$\mu$m, PACS 70~$\mu$m, and PACS 100~$\mu$m. Blue pixels have a low flux density whereas red pixels have a high flux density, following an arcsinh scale. The colours used are simply chosen to best represent the large dynamical range of intensities across all bands and all pixel sizes and should be used in a qualitative sense only.\label{fig:maps-convolved}}
\end{figure}

Because pixel--by--pixel analysis will be central for this study, it is crucial to match all the images to a common reference frame. To do so, it is important that all bands share a common point spread function (PSF). To ensure this, in a first step we have convolved all the images to the PACS 100~$\mu$m PSF using the dedicated kernels provided by \cite{aniano2011a}. We have then registered these images to a common reference frame with a pixel size ranging from 8\arcsec, slightly larger than the PACS 100~$\mu$m PSF, to 512\arcsec, by increments of 1\arcsec\ in terms of pixel size, using \textsc{iraf}'s \textsc{wregister} procedure with the \textsc{drizzle} interpolant. This allows us to sample all scales from fractions of H\,\textsc{ii} regions at 33~pc (8\arcsec) to large fractions of the disk at 2084~pc (512\arcsec). The upper bound is limited by the size of the galaxy. Increasing to larger physical scales would leave us with too few pixels in M33. We present some of the final, convolved, registered, and background subtracted maps used in this study for a broad range of pixel sizes in Fig.~\ref{fig:maps-convolved}.

To compute flux uncertainties, we have relied on the 33~pc scale maps. We took into account the uncertainty on the background determination, which is due to large scale variations, and the pixel--to--pixel noise. The former was measured as the standard deviation of the background level measured within $10\times10$ pixel square apertures around the galaxy using \textsc{iraf}'s \textsc{imexamine} procedure. The latter was measured as the mean of the standard deviation of pixel fluxes in these apertures around the galaxy. We then summed these uncertainties in quadrature. For maps at lower resolution, we simply scaled the uncertainties on the background with the square of the pixel size, and the pixel--to--pixel uncertainties with the pixel size. Direct measurements on lower resolution maps yielded uncertainties consistent with the scaled ones.

\subsection{Removal of the stellar pollution in infrared bands}

In a final data processing step, we have removed the stellar contamination in the 8~$\mu$m and 24~$\mu$m bands. To do so, we have assumed that the \textit{Spitzer}/IRAC 3.6~$\mu$m image is dominated by stellar emission, following the analysis of \cite{meidt2012a}. We have then scaled this image to predict the stellar emission at 8~$\mu$m and 24~$\mu$m and subtracted it from these images. We have assumed a scaling factor of 0.232 at 8~$\mu$m and 0.032 at 24~$\mu$m, following \cite{helou2004a}. We should note that this scaling factor can change quite significantly with the star formation history \citep[e.g.][]{calapa2014a,ciesla2014a}.

\section{Comparison of SFR estimators at different scales\label{sec:comp-SFR-scales}}

\subsection{Presentation of monochromatic and composite SFR estimators\label{ssec:SFR-estimators}}

Ideally, a good SFR estimator has a solid physical basis and is devoid of biases. Thus, because they trace directly or indirectly the emission from young, massive stars, the attenuation\footnote{We distinguish between the extinction which includes the absorption and the scattering out of the line of sight, and the attenuation which also includes the scattering into the line of sight. In practice in this study we only have access to the attenuation and not to the extinction. See for instance Sect. 1.4.1 of \cite{calzetti2013a}.}--corrected FUV or H$\alpha$ should in principle be ideal estimators. In practice however, the presence of biases is a real problem since it shows that other factors unrelated to recent star formation can contribute to the emission in star--formation tracing bands. For instance, in the case of monochromatic IR tracers, such factors are the contribution from old stars, changes in the opacity of the ISM (interstellar medium), or in the IR SED (spectral energy distribution).

When no attenuation measurement is available, a popular method developed over the last few years has been to combine attenuated and attenuation free tracers \citep{calzetti2007a,kennicutt2007a,leroy2008a,kennicutt2009a,hao2011a}. Unfortunately, how to combine such tracers remains uncertain. \cite{calzetti2007a} and \cite{kennicutt2009a} found different scaling factors when combining dust emission at 24~$\mu$m with H$\alpha$, likely because of different scales probed: 500~pc for the former and entire galaxies for the latter, and therefore different timescales \citep{calzetti2013a}. According to \cite{leroy2012a}, the universality of composite estimators remains in doubt. One of the main issues comes from the diffuse emission and whether it is linked to recent star formation or not. In M33, the fraction of diffuse emission is high: 65\% in FUV, and from 60\% to 80\%  in the 8~$\mu$m and 24~$\mu$m bands, with clear variations across the disk for the latter two \citep{verley2009a}. While some methods have been suggested to remove the diffuse emission linked to old stars \citep{leroy2012a}, they rely on uncertain assumptions. We therefore cannot rely a priori on such tracers as an absolute reference. Yet, how monochromatic and hybrid SFR estimators compare may still yield useful information on star formation in M33. We consider the restricted set of monochromatic and hybrid SFR estimators presented in Table~\ref{tab:SFR}.

\begin{table}
\caption{SFR estimators.}
\label{tab:SFR}
\centering
\begin{tabular}{c c c c c c c c c}
 \hline\hline
\multicolumn{5}{c}{Monochromatic}\\\hline
Band&log C$_\mathrm{band}$&k&Method&Reference\\\hline
FUV&$-36.355$&1.0000&Theoretical\textsuperscript{a}&1\\
H$\alpha$&$-34.270$&1.0000&Theoretical\textsuperscript{a}&1\\
24~$\mu$m&$-29.134$&0.8104&H$\alpha$\textsuperscript{b}&2\\
70~$\mu$m&$-29.274$&0.8117&H$\alpha$\textsuperscript{b}&3\\
100~$\mu$m&$-37.370$&1.0384&H$\alpha$\textsuperscript{b}&3\\\hline\hline
\multicolumn{5}{c}{Hybrid}\\\hline
Band&log C$_\mathrm{band1}$&$\mathrm{k_{band1-band2}}$&Method&Reference\\\hline
H$\alpha$+24~$\mu$m&$-34.270$&$0.031$&H$\alpha$\textsuperscript{b}&2\\
FUV+24~$\mu$m&$-36.355$&$6.175$&H$\alpha$+24~$\mu$m&4\\
\hline
\end{tabular}
\tablebib{(1) \cite{murphy2011a}, (2) \cite{calzetti2007a}, (3) \cite{li2013a}, (4) \cite{leroy2008a}}
\tablefoot{Monochromatic: $\mathrm{\log \Sigma SFR=\log C_{band}+k\times\log S_{band}}$; Hybrid: $\mathrm{\log \Sigma SFR=\log C_{band1}+\log\left[S_{band1}+k_{band1-band2}\times S_{band2}\right]}$, with $\Sigma$SFR in M$_\odot$~yr$^{-1}$~kpc$^{-2}$, S defined as $\mathrm{\nu S_\nu}$ in W~kpc$^{-2}$, and C in M$_\odot$~yr$^{-1}$~W$^{-1}$.} Empirical estimators have been calibrated on individual star--forming regions on typical scales of the order of $\sim$200--500~pc.\\
\textsuperscript{a} Based on Starburst99 \citep{leitherer1999a}.
\textsuperscript{b} Extinction corrected, calibrated against near-infrared hydrogen recombinations lines (e.g., Pa$\alpha$ or Br$\gamma$).
\end{table}

Before comparing these SFR estimators we have to add a word of caution. In some cases, especially at the smallest spatial scales, the concept of an SFR in itself may not be valid \citep[for a description of the reasons see Sect. 3.9 of][in particular: IMF sampling, age effects, and the spatial extension of the emission in star--formation--tracing bands in comparison to the resolution]{kennicutt2012a}. In the context of this study, IMF sampling is not likely to be a particular issue. A scale of 33~pc corresponds to the Str\"omgren radius of a 3000--5000~M$_\odot$, 4--5~Myr old stellar cluster. Such a cluster would already be massive enough not to be too affected by stochastic sampling \citep{fouesneau2012a}. However, we cannot necessarily assume that other assumptions are fulfilled: age effects may be strong and star--forming regions may be individually resolved \citep{kennicutt2012a,kruijssen2014a}. This means that care must be taken when interpreting the SFR. In this case it may be preferable to interpret the SFR as a proxy for the local radiation field intensity. The dust emission may come from heating by local old stellar populations or because of heating by energic photons emitted by stars in neighbouring pixels rather than being driven by local massive stars.

\subsection{Comparison between monochromatic and hybrid SFR estimators}

\begin{figure*}[!htbp]
 \includegraphics[width=\columnwidth]{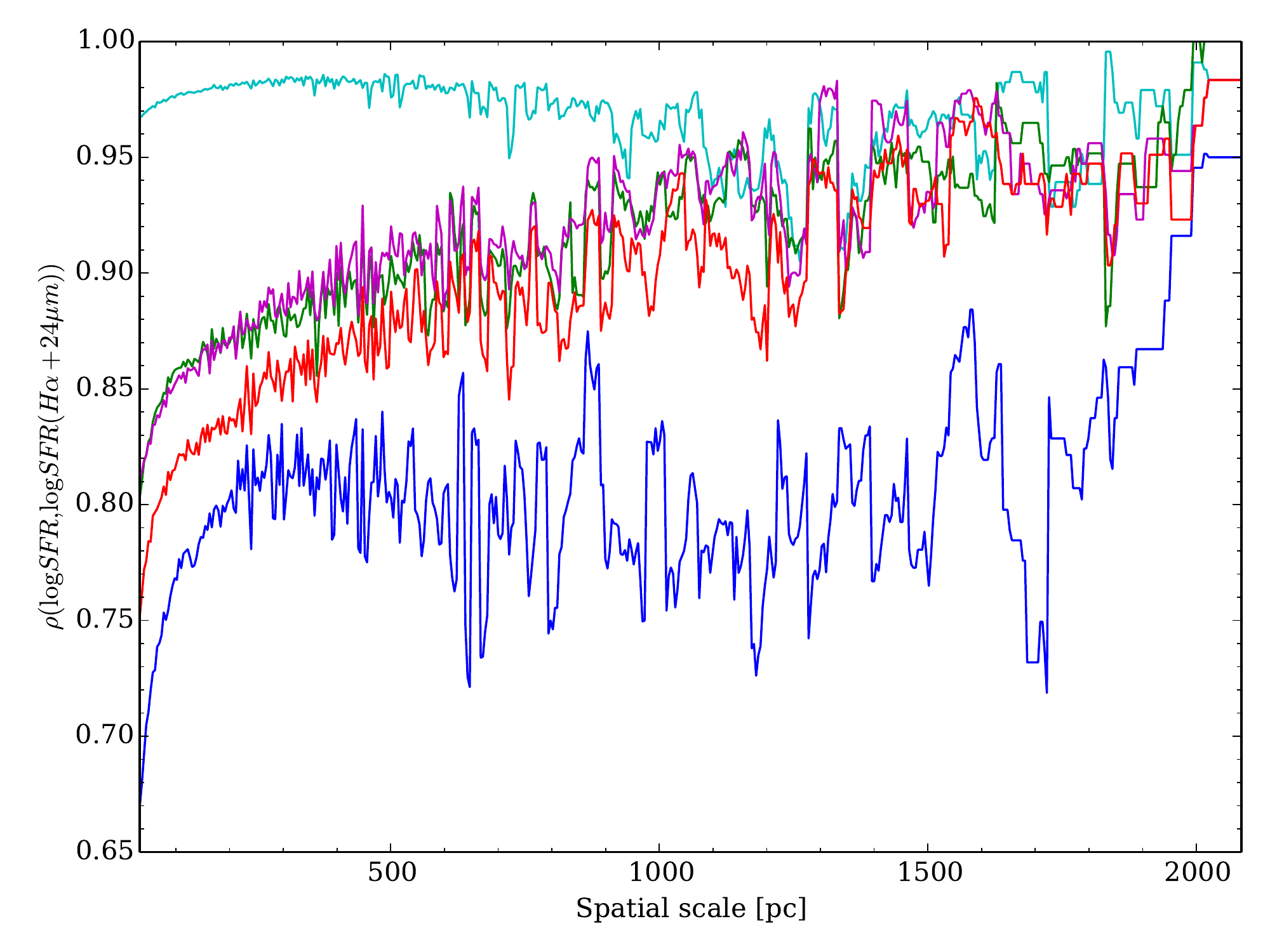}
 \includegraphics[width=\columnwidth]{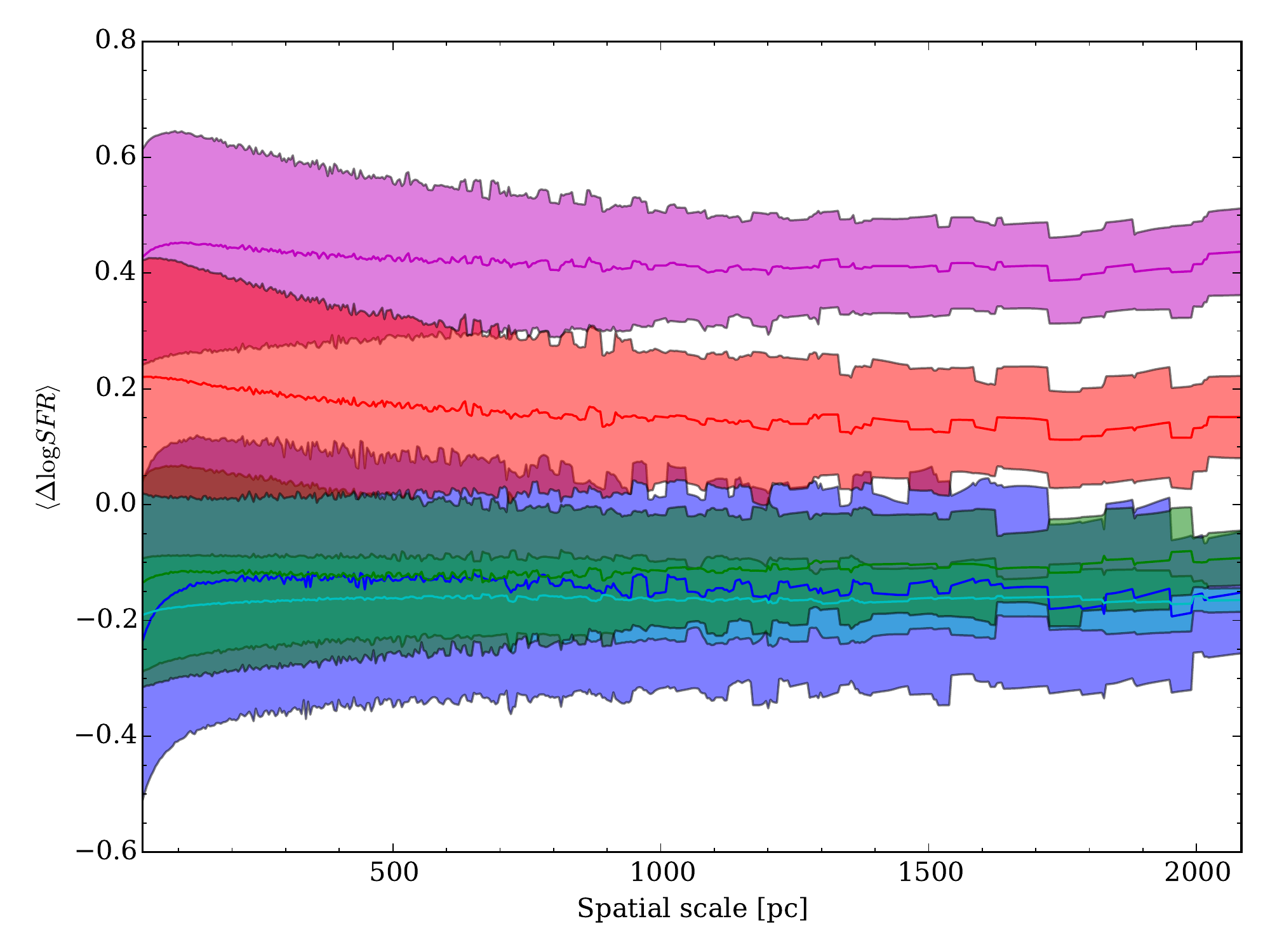}
 \caption{Comparison of monochromatic SFR estimators with the reference SFR(H$\alpha$+24~$\mu$m) estimator versus the spatial scale. The colour indicates the monochromatic band: FUV (blue), H$\alpha$ (cyan), 24~$\mu$m (green), 70~$\mu$m (magenta), and 100~$\mu$m (red). The left panel shows the correlation coefficient between the monochromatic and reference estimators. The right panel shows the mean offset (solid line) and the standard deviation (shaded area) of the difference between the estimators.
 \label{fig:comp-sfr}}
\end{figure*}

We now compare popular monochromatic SFR estimators in the FUV, H$\alpha$ (both uncorrected for the attenuation), 24~$\mu$m, 70~$\mu$m, and 100~$\mu$m bands with SFR(H$\alpha$+24~$\mu$m), which we take as the refererence, to understand how their relation changes with the scale considered. We have selected SFR(H$\alpha$+24~$\mu$m) over SFR(FUV+24~$\mu$m) as we will see in Sect.~\ref{sec:discussion}, at local scales the 24~$\mu$m and the H$\alpha$ are more closely related. We should note that in this study we are not so much interested in the absolute SFR, which we cannot compute reliably at all scale, as in the consistency of SFR estimators with one another and their relative variations with spatial scale. These relative variations bring us important information on star formation in M33. In addition, if different estimators give systematically different results, this shows that they cannot all be simultaneously reliable. The relations between the various aforementioned SFR estimators are shown in Fig.~\ref{fig:comp-sfr}.

We first observe that monochromatic SFR estimates are well correlated with the reference estimates ($0.67\le\rho\le1.00$, with $\rho$ the Spearman correlation coefficient). There is a rapid increase of the correlation coefficient up to a scale of 150--200~pc for all estimators but H$\alpha$. Beyond 200~pc, IR estimators show a regular increase. The FUV correlation coefficient remains relatively stable until a scale of 1700~pc and then rapidly increases. The H$\alpha$ estimator globally shows little variation with scale. At scales beyond 2~kpc, all estimators are strongly correlated with the reference one.

However, if they are all well correlated, this does not necessarily mean that they provide consistent results. In the right panel of Fig.~\ref{fig:comp-sfr}, we show the mean offset and the dispersion between monochromatic SFR estimators and the reference one. The FUV, H$\alpha$, and 24~$\mu$m estimates are lower than the reference one. This is naturally expected for H$\alpha$ because it is part of the reference SFR estimator. The FUV being subject to the attenuation will also naturally yield lower estimates. The amplitude of the offset at 24~$\mu$m (0.14~dex at 33~pc to 0.10~dex at 2084~pc) can be more surprising as the 24~$\mu$m estimator used here is non--linear to take into account that only a fraction of photons are attenuated by dust. This is probably due to a metallicity effect. \cite{magrini2009a} measured the metallicity of M33 H\,\textsc{ii} regions at $12+\log O/H=8.3$, placing it near the limit between the high ($12+\log O/H>8.35$) and intermediate ($8.00<12+\log O/H\le8.35$) metallicity samples of \cite{calzetti2007a}. In turn, intermediate metallicity galaxies show some deficiency in their 24~$\mu$m emission relative to higher metallicity galaxies. This is due to reduced dust content of the ISM which increases its transparency \citep{calzetti2007a}.

If we compare the SFR at 24~$\mu$m and 70~$\mu$m, the relative offset ranges from $0.57$~dex at 33~pc to $0.53$~dex at 2084~pc. Such a discrepancy has several possible origins. First, these infrared estimators have been determined only for a limited range in terms of $\Sigma$SFR. \cite{li2013a} computed their estimators for $\mathrm{-1.5\leq\log\Sigma SFR\leq0.5~M_\odot~yr^{-1}~kpc^{-2}}$. The 24~$\mu$m estimator of \cite{calzetti2007a} benefited from a much broader range: $\mathrm{-3.0\leq\log\Sigma SFR\leq 1.0~M_\odot~yr^{-1}~kpc^{-2}}$. If we consider only the definition range of SFR(70~$\mu$m), at the finest pixel size, the disagreement between SFR(70~$\mu$m) and SFR(H$\alpha$+24~$\mu$m) is not as strong. Another possible source of disagreement lies in the scale at which estimators have been derived. Indeed, increasing the pixel sizes means averaging over larger regions and including a larger fraction of diffuse emission. \cite{li2013a} determined their estimators on two galaxies at a scale of about 200~pc. They estimated that 50\% of the emission at this scale comes from dust heated by stellar populations unrelated to the latest episode of star formation. Yet even if this diffuse emission were exclusive to the 70~$\mu$m band, this would not be sufficient to explain the full extent of the offset. \cite{calzetti2007a} combined data of a much more diverse sample of galaxies at a physical scale ranging from 30~pc to 1.26~kpc, averaging out specificities of individual galaxies. To gain further insight on these differences, we will examine in detail the origin of dust emission at different scales in Sect.~\ref{sec:scale-dep-estimators}.

As a concluding remark, these discrepancies must serve as a warning when using SFR estimators. Their application beyond their validity range in terms of surface brightness, physical scale, metallicity may yield important biases. This is especially important when applying SFR estimators on higher redshift galaxies as their physical properties may be more poorly known.

\section{Understanding dust emission to measure the SFR at different scales\label{sec:scale-dep-estimators}}

\subsection{What the infrared emission traces from 24~$\mu$m to 100~$\mu$m}

To understand what the emission of the dust traces at which scale and under which conditions, we examine the change in the relative emission at 24~$\mu$m, 70~$\mu$m, and 100~$\mu$m. To ease the comparison, we first convert the luminosity surface densities into SFR using the linear estimators of \cite{rieke2009a} at 24~$\mu$m and \cite{li2013a} at 70~$\mu$m and 100~$\mu$m. We stress that we are not interested here in the absolute values of SFR but only their relative variations. These linear estimators only serve to put luminosities on a comparable scale.

\begin{figure*}[!htbp]
\includegraphics[width=\textwidth]{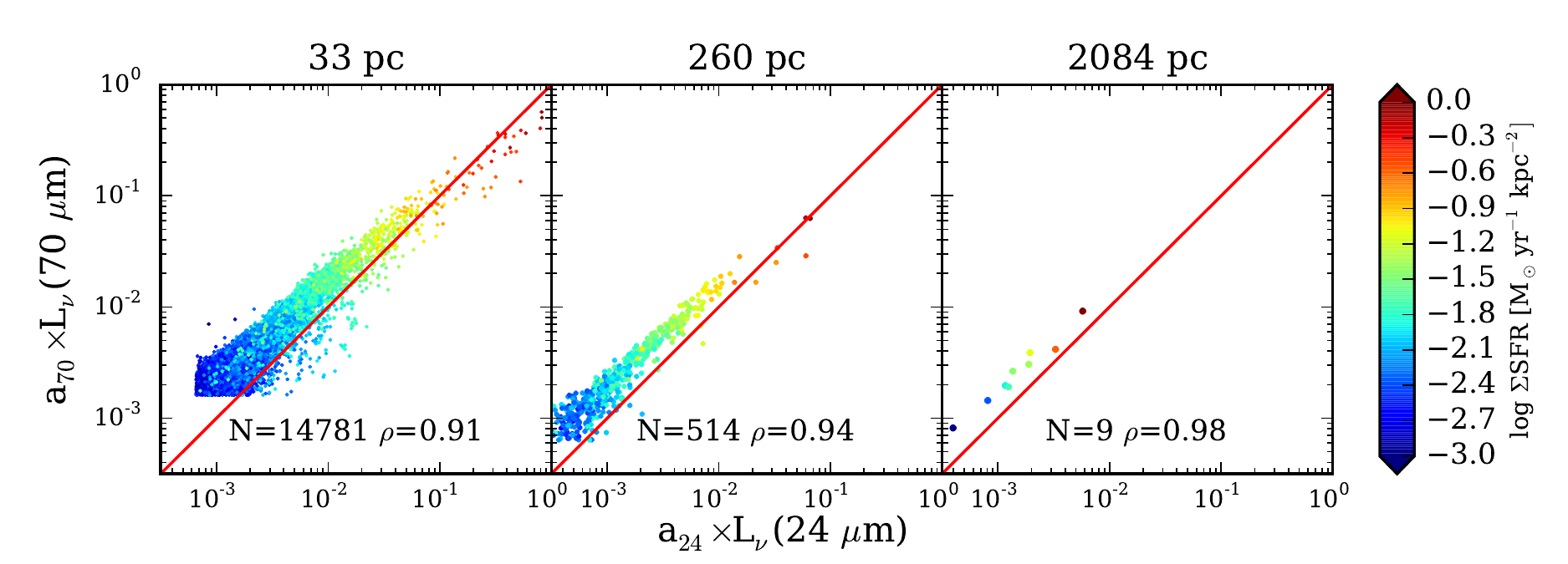}
\includegraphics[width=\textwidth]{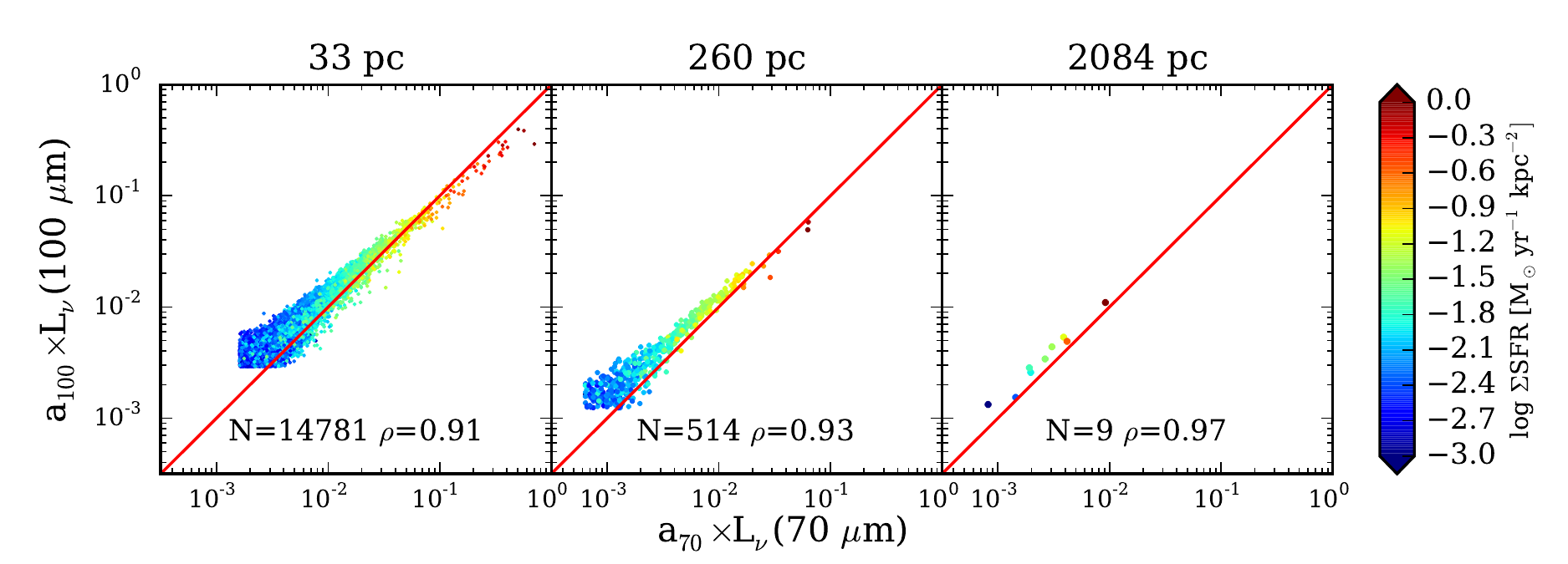}
\caption{Relations at 33~pc (left), 260~pc (centre), and 2084~pc (right) of L$_\nu$(70~$\mu$m) versus L$_\nu$(24~$\mu$m) (top), and  L$_\nu$(100~$\mu$m) versus L$_\nu$(70~$\mu$m) (bottom). All luminosities have been multiplied by a constant factor corresponding to a linear SFR estimator (a$_{24}=2.04\times10^{-36}$~M$_\odot$~yr$^{-1}$~W$^{-1}$, a$_{70}=5.89\times10^{-37}$~M$_\odot$~yr$^{-1}$~W$^{-1}$, and a$_{100}=5.17\times10^{-37}$~M$_\odot$~yr$^{-1}$~W$^{-1}$) in order to put them on a similar scale. The colour of each point indicates $\Sigma$SFR following the colour bar at the right of each row. The red line indicates a one--to--one relation. We see the non--linear relations between the luminosities in different bands. These non--linearities are particularly apparent at the finest pixel scales. At coarser scales, the relations appear more linear, which is probably due to a mixing between diffuse and star--forming regions.\label{fig:comp-sfr-24-70-100}}
\end{figure*}

In Fig.~\ref{fig:comp-sfr-24-70-100} we compare the dust emission at 24~$\mu$m, 70~$\mu$m, and 100~$\mu$m from 33~pc to 2084~pc.
In general there is an excellent correlation between the emission in these three bands across all scales ($0.90\leq\rho\leq0.98$). Unsurprisingly the luminosity of individual regions in all bands also varies with $\Sigma$SFR. When examining relations at a scale of 33~pc, we find that there is a systematic sub--linear trend between shorter and longer wavelength bands. For higher luminosity surface densities, L$_\nu$(24~$\mu$m) is stronger relatively to L$_\nu$(70~$\mu$m) than what can be seen at lower L$_\nu$(24~$\mu$m) or L$_\nu$(70~$\mu$m). The same behaviour is clearly observed when comparing L$_\nu$(70~$\mu$m) with L$_\nu$(100~$\mu$m). Interestingly, when going towards coarser resolutions this trend progressively disappears, and at 2084~pc the relations between the various bands appear more linear. The important aspect to note is not so much that the dispersion diminishes with coarser spatial scales but that there is a progressive transition from a non--linear relation to a linear relation. This phenomenon could be due to the progressive mixing of diffuse and star--forming regions.

\begin{figure*}[!htbp]
\includegraphics[width=\columnwidth]{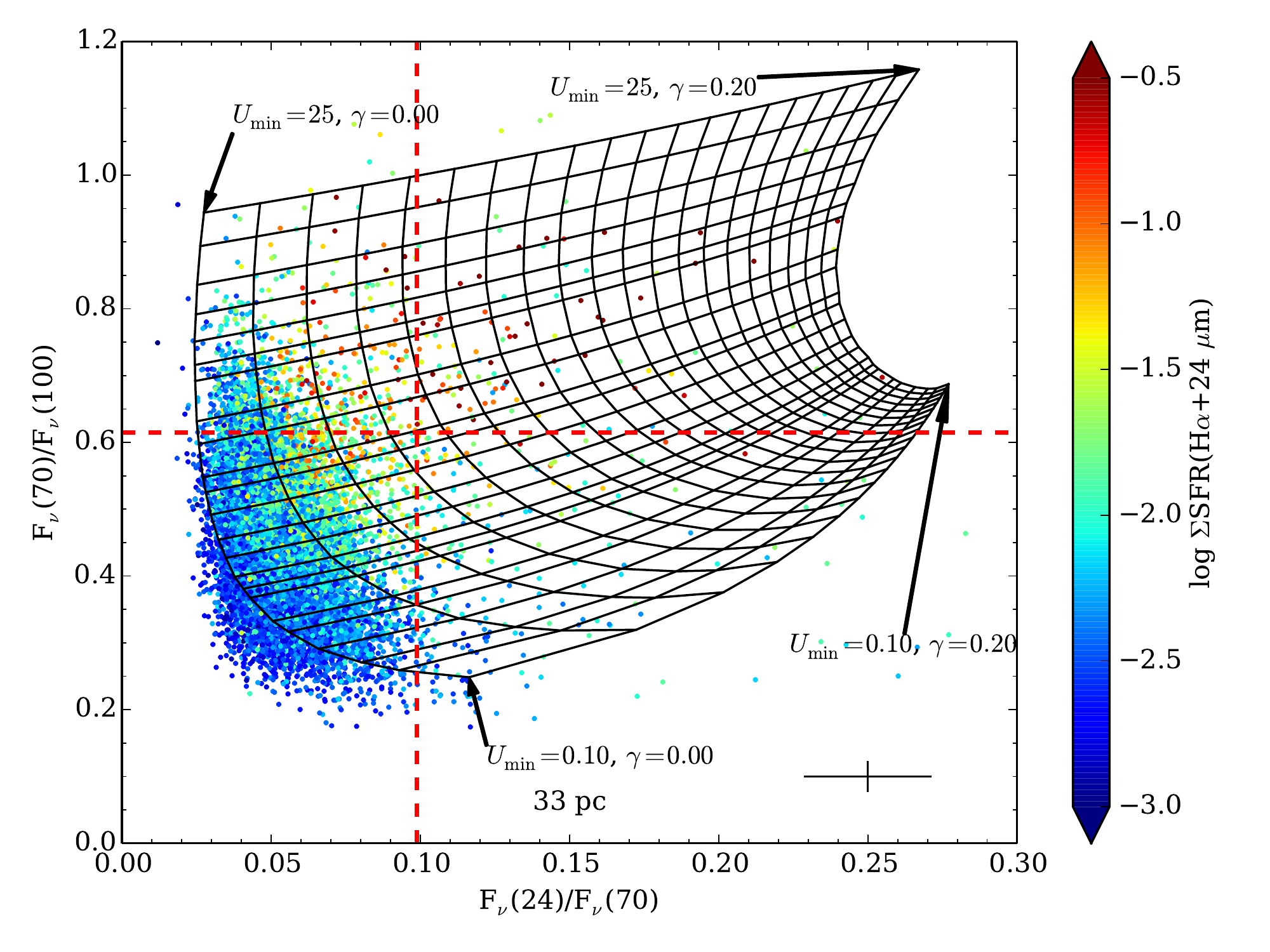}
\includegraphics[width=\columnwidth]{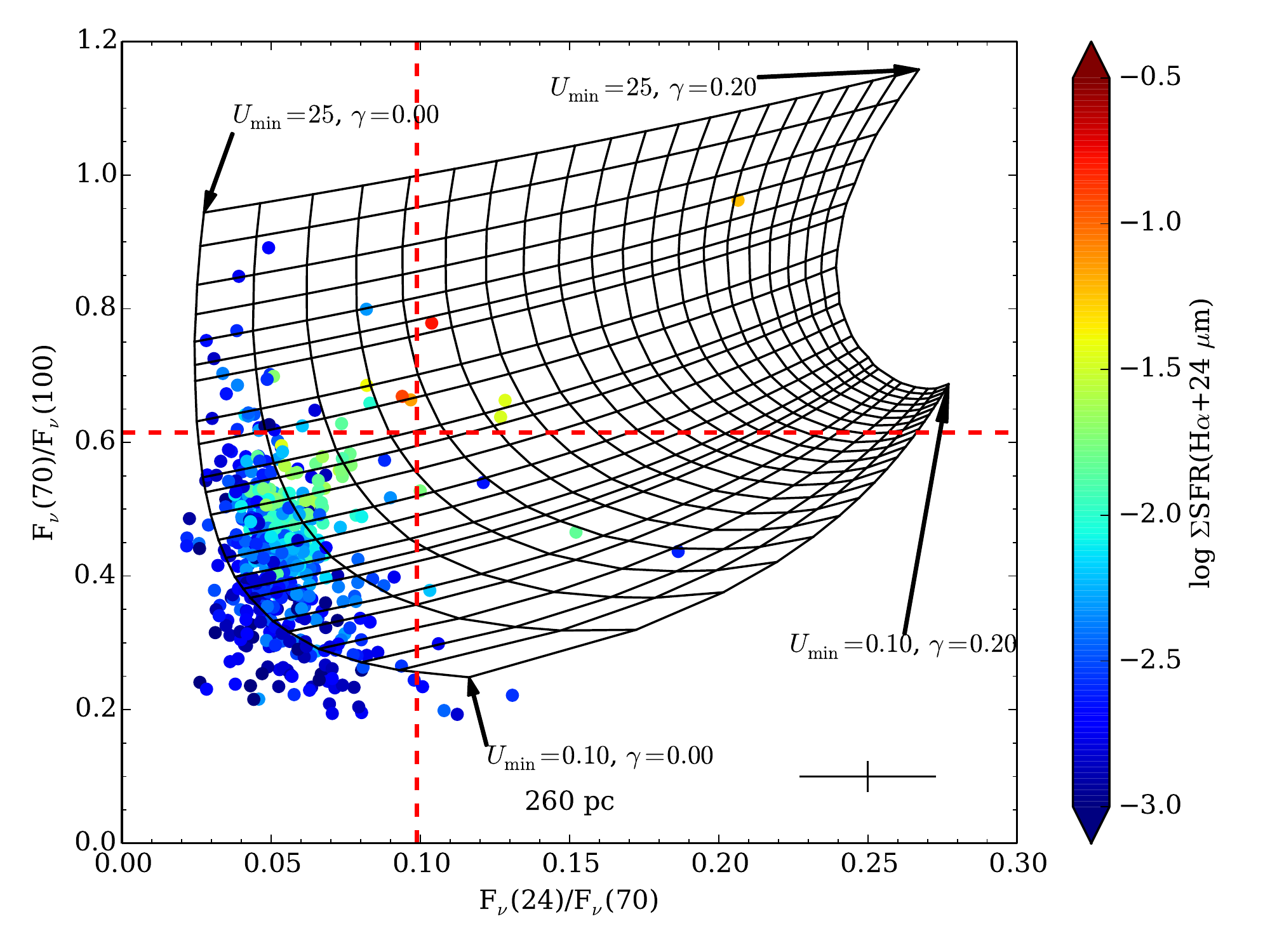}
\caption{70--to-100 versus 24--to--70 flux density ratios for each pixel at a resolution of 33~pc (left) and 260~pc (right). The colour of each symbol corresponds to $\Sigma$SFR, according to the colour bar on the right. The grid represents the \cite{draine2007a} models, with the MW3.1\_30 dust composition, $0.10\le U_\textrm{min}\le 25$, and $0.00\le\gamma\le0.20$. The red dashed lines indicate the locus corresponding to the one--to--one relations shown in Fig.~\ref{fig:comp-sfr-24-70-100}. The 3--$\sigma$ uncertainties are shown in the bottom right corner. We see that the 24--to--70 ratio is well correlated with $\gamma$ and $\Sigma$SFR, especially at 33~pc. At 260~pc, due to mixing between diffuse and star--forming regions, excursions in $\gamma$ are strongly reduced. Note that when considering the galaxy as a whole, a large fraction of the emission is due to the handful of luminous regions rather than the larger number of faint regions.\label{fig:comp-draine-24-70-100}}
\end{figure*}

To understand how the relative infrared emission varies with the spatial scale, we compare the observed dust at 24~$\mu$m, 70~$\mu$m, and 100~$\mu$m with the model of \cite{draine2007a}. We refer to Rosolowsky et al. (in prep.) for a full description of the dust SED modelling of M33 with the \cite{draine2007a} models. In a nutshell, the emission of the dust is modelled by combining 2 components. The first component is illuminated by a starlight intensity $U_\textrm{min}$, corresponding to the diffuse emission. The other component corresponds to dust in star forming regions, illuminated with a starlight intensity ranging from $U_\textrm{min}$ to $U_\textrm{max}$ following a power law. We consider all available values for $U_\textrm{min}$, from  0.10 to 25. Following \cite{draine2007b}, we adopt a fixed $U_\textrm{max}=10^6$. The fraction of the dust mass linked to star--forming regions is $\gamma$, and as a consequence $1-\gamma$ is the mass fraction of the diffuse component. We consider $\gamma$ ranging from 0.00 to 0.20 by steps of 0.01. Because M33 has a sub--solar metallicity, we adopt the so--called MW3.1\_30 dust composition, which corresponds to a Milky Way dust mix with a PAH mass fraction relative to the total dust mass of 2.50\%, lower than the Milky Way mass fraction of 4.58\%. We compare this grid of physical models to the observations in Fig.~\ref{fig:comp-draine-24-70-100} for a resolution of 33~pc, and at a resolution of 260~pc.

We see that the parameters space spanned by the grid of models reproduces the observations very well except for a fraction of points at low 24--to--70 and 70--to--100 ratios, for which even models with $\gamma=0$ fail. Most points are concentrated in regions with simultaneously low values for $\gamma$ and $U_\textrm{min}$, which also correspond to low SFR estimates. Regions at higher SFR seem to have a higher value for $U_\textrm{min}$ and there is a clear trend with $\gamma$, strongly star--forming regions having a larger $\gamma$. In other words, this means that the relative increase of the 24~$\mu$m emission compared to the 70~$\mu$m one that we saw in Fig.~\ref{fig:comp-sfr-24-70-100} is likely due to the transition between a regime entirely driven by the diffuse emission and a nearly complete lack of dust heated in star--forming regions ($0.00\le\gamma\le0.01$), to a regime with a strong contribution from dust heated in star--forming regions. When the resolution is coarser the emission from star--forming regions is increasingly mixed with the emission from dust illuminated by the diffuse radiation field, reducing the excursions to large values of $\gamma$ required to have a strong emission at 24~$\mu$m compared to the emission at 70~$\mu$m. If we assume that on average in star-forming galaxies $\gamma=1-2$\% \citep[e.g.][]{draine2007b}, a significant fraction of the luminosity at 70~$\mu$m comes from star--forming regions. Considering a resolution of 33~pc, these values of $\gamma$ correspond typically to regions with $\log \Sigma \text{SFR}\ge-2$ to $-1.5$~M$_\odot$~yr$^{-1}$~kpc$^{-2}$. The 70~$\mu$m luminosity contributed by regions brighter than $\log \Sigma \text{SFR}=-2$ and -1.5 is 58\% and 27\% respectively. This is consistent with what we would expect from Fig.~\ref{fig:comp-draine-24-70-100} as a small fraction of pixels with a high $\Sigma \text{SFR}$ contributes a large fraction to the total luminosity compared to the more numerous but much fainter pixels.

We can also understand the observed trends by examining the physical origin of dust emission in relation to the SFR. At high SFR, the emission at 24~$\mu$m and 70~$\mu$m is caused by dust at the equilibrium and by a stochastically heated component. In low SFR regions only the stochastically heated component remains at 24~$\mu$m, contrary to what occurs at 70~$\mu$m \citep[see in particular Fig. 15 in][]{draine2007a}. This means that the 24~$\mu$m emission should drop more quickly than the 70~$\mu$m emission with decreasing SFR. This accounts for the difference in behaviour seen in Fig.~\ref{fig:comp-draine-24-70-100}. The preceding explanation for M33 seems consistent with the findings of \cite{calzetti2007a,calzetti2010a} who have studied this problem in great detail. Combining several samples totalling almost 200 star--forming galaxies, \cite{calzetti2010a} also found a clear positive correlation between the measured SFR and the 24--to--70 ratio.
\subsection{Impact of the scale on the measure of the SFR from monochromatic infrared bands}

\subsubsection{Computation of SFR scaling relations}

The determination of the SFR is paramount to understanding galaxy formation and evolution. Initially, such estimates in the mid-- and far--infrared were limited to entire galaxies due to the coarse resolution of the first generations of space--based IR instruments. \textit{Spitzer} has enabled the computation of dust emission in galaxies at a local scale in nearby galaxies \citep[e.g.,][]{boquien2010a}. Thanks to its outstanding resolution, \textit{Herschel} has enabled such studies at the peak of the emission of the dust at ever smaller spatial scales \citep{boquien2010b,boquien2011a,galametz2013a}. But such a broad and homogeneous spectral sampling represents an ideal case. More commonly, just one or a handful of infrared bands are available at sufficient spatial resolution. It is therefore important not only to be able to estimate the SFR from just one or few IR bands but also to understand how this is dependent on the spatial scale.

To do so, we simply determine at each resolution the scaling factor $\mathrm{C_{band}}$ between a given band and $\Sigma$SFR from the combination of H$\alpha$ and 24~$\mu$m. This is done by carrying out an orthogonal distance regression using the \textsc{odr} module from the \textsc{scipy} \textsc{python} library on the following relation:
\begin{equation}
 \mathrm{\log\Sigma SFR=\log{C_{band}}+\log S_{band}}.\label{eqn:fit}
\end{equation}
In order to examine the difference between intense and quiescent regions, at each resolution we have also separated the regions into 4 bins in addition to fitting the complete sample: top and bottom 50\%, and top and bottom 15\%, in terms of $\Sigma$SFR from the combination of H$\alpha$ and 24~$\mu$m. The most extreme bins ensure that we only select the most star--forming (top) or the most diffuse (bottom) regions in the galaxy.

\subsubsection{Dependence of SFR scaling relations on the pixel size}

\begin{figure*}[!htbp]
\includegraphics[width=\textwidth]{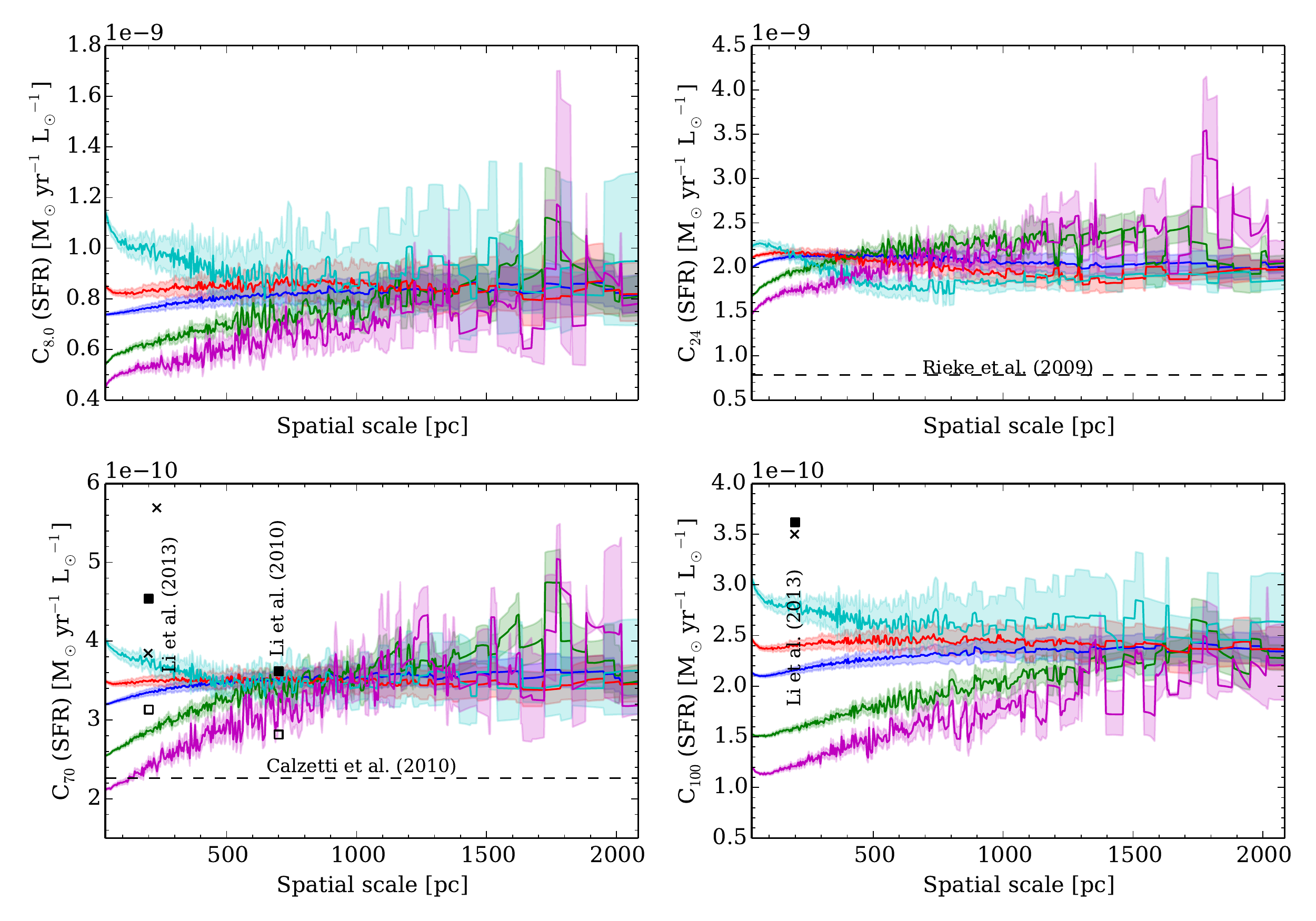}
\caption{Scaling coefficients from the luminosity in infrared bands to $\Sigma$SFR versus the pixel size, at 8~$\mu$m, 24~$\mu$m, 70~$\mu$m, and 100~$\mu$m, from the top left corner to the bottom right corner. The blue line indicates the value of the scaling factor when taking into account all pixels detected at a 3--$\sigma$ level in all six bands. The red (respectively green) line indicates the scaling factor when considering only pixels with a $\Sigma$SFR higher (resp. lower) than the median $\Sigma$SFR at a given resolution. The cyan and magenta lines represent regions in the top and bottom 15\% in terms of $\Sigma$SFR. The shaded areas of the corresponding colours indicate the 1--$\sigma$ uncertainties. The horizontal dashed line at 24~$\mu$m (resp. 70~$\mu$m) indicates the scaling factor determined by \cite{rieke2009a} \citep[resp. ][]{calzetti2010a} for entire galaxies. The crosses for the 70~$\mu$m and 100~$\mu$m bands indicate the scaling factor determined for individual galaxies at a scale of 200~pc \citep{li2013a} and 700~pc \citep{li2010a}. The squares indicate mean values over several galaxies. The empty squares denote that no background subtraction was performed.\label{fig:sfr-calib-lin}}
\end{figure*}

The dependence of the scaling factors on resolution at 8~$\mu$m, 24~$\mu$m, 70~$\mu$m, and 100~$\mu$m is presented in Fig.~\ref{fig:sfr-calib-lin}.
\paragraph{Description of the scaling relations}It clearly appears that regions with strong and weak $\Sigma$SFR have markedly different scaling factors and a different evolution with pixel size. Compared to the entire sample, at 33~pc the scaling factor for the 50\% (respectively 15\%) brightest pixels is higher by a factor 1.06 to 1.16 (resp. 1.12 to 1.55). Conversely, the scaling factor for the 50\% (resp. 15\%) faintest pixels is lower by a factor 0.71 to 0.84 (resp. 0.56 to 0.74). When increasing the pixel size from 33~pc to 2084~pc, the scaling factor for pixels with a weak $\Sigma$SFR strongly increases. On the opposite, the scaling factor for pixels with a strong $\Sigma$SFR generally shows a slightly decreasing trend. From a typical scale of 400~pc to 1200~pc, depending on the infrared band, there is no significant difference in the scaling factors between pixels with weak and strong $\Sigma$SFR.

\paragraph{Impact of the relative fraction of diffuse emission} As we have already explained, our reference SFR estimator combining H$\alpha$ and 24~$\mu$m is unfortunately not perfect because it is also sensitive to diffuse emission that may or may not actually be related to star formation. We now consider only the 15\% brightest pixels at 33~pc. They most likely correspond to pure star forming regions with little or no diffuse emission. Conversely, the 15\% faintest pixels will be almost exclusively made of diffuse emission with little or no local star formation. That way the scaling factor will be higher for the former compared to the latter. If we move to coarser resolutions, individual pixels will increasingly be made of a mix of star--forming and diffuse regions such that the brightest and faintest regions will be less different at 2084~pc than they are at 33~pc. This naturally yields increasingly similar scaling factors that progressively lose their dependence on the intensity of star formation. In other words, this means that on a scale larger than roughly 1~kpc, monochromatic IR bands from 8~$\mu$m to 100~$\mu$m may be as reliable for estimating the SFR as the combination of H$\alpha$ and 24~$\mu$m. This scale is probably indicative of the typical scale from which there is always a similar fraction of diffuse and star--forming regions in each pixel, bright or faint. This scale is likely to vary depending on intensity of star formation in a given galaxy and on its physical propeties. This aspect should be explored in a broader sample of spiral galaxies. We also have to mention that this result is also affected by the transparency of the ISM as we will see below, or by non--linearities that are not accounted for here. For instance, in intense star--forming regions the 8~$\mu$m emission may get depressed because of the PAH destruction by the strong radiation field \citep{boselli2004a,helou2004a,bendo2006a}, or because the 8~$\mu$m has a strong stochastic component, proportionally more important than at 24~$\mu$m. These processes can induce a non proportionality between the 8~$\mu$m emission and the SFR. Finally, we note that the difference in the scaling factor between the faintest and the brightest bins is minimal at 24~$\mu$m. This is most likely due to the fact that the 24~$\mu$m emission affects both sides of Eq.~\ref{eqn:fit}. 

\paragraph{Comparison with the literature} When comparing the scaling factors determined in M33 with those determined in the literature from both individual star--forming regions in galaxies and entire galaxies, we find instructive discrepancies. At a scale of 200~pc, the scaling factors at 70~$\mu$m determined by \cite{li2013a} for NGC~5055 and NGC~6946 are systematically higher. As discussed in the aforementioned article, this may be due to background subtraction. Indeed their study is based on the selection of individual H\,\textsc{ii} regions, allowing for the subtraction of the local background, which is not easily doable with accuracy when carrying out a systematic pixel--by--pixel analysis like we are doing in this article. Without background subtraction, they obtain a scaling factor that is very similar to the one we find when selecting pixels with a strong SFR. A similar study carried out at a scale of 700~pc by \cite{li2010a} leads to a similar result.

When we compare our scaling factors to the ones obtained on entire galaxies at 24~$\mu$m by \cite{rieke2009a} and at 70~$\mu$m by \cite{calzetti2010a} there is a clear discrepancy, their scaling factors being lower. Because we see little trend with pixel size at larger scales, it appears unlikely that the scaling factor will diminish strongly at scales larger than 2084~pc. A possible explanation is that this could be due to the increased ISM transparency in M33. In other words, this could be because a smaller fraction of the energetic radiation emitted by young stars is reprocessed by dust into the infrared. In the case of M33, about 75\% of star formation is seen in H$\alpha$ and only 25\% in the infrared. Indeed, \cite{li2010a} found a trend of the scaling factor with the oxygen abundance, with more metal--poor galaxies having a higher coefficient. If we consider the relation \cite{li2010a} find between the oxygen abundance and the scaling factor, the change in the coefficient from $12+\log O/H\simeq8.3$ (for M33) to $12+\log O/H\simeq8.7$ \citep[for the sample of ][]{calzetti2010a}, would explain the observed discrepancy. At the same time, we notice that the discrepancy with \cite{rieke2009a} at 24~$\mu$m is more important than with \cite{calzetti2010a} at 70~$\mu$m. This is expected because the former sample is made of the most deeply dust--embedded galaxies ([ultra] luminous infrared galaxies), contrary to the latter one which is made of galaxies that are more transparent at short wavelength. We should however note that at a given metallicity, \cite{li2010a} find an important dispersion. This is exemplified by the case of NGC~5055 and NGC~6946, which despite having very similar metallicities yield very different scaling factors. In addition, a galaxy like the Large Magellanic Cloud which has a metallicity similar to that of M33 has a scaling factor similar to that of galaxies with $12+\log O/H\simeq8.7$, perhaps because it has been calibrated with HII regions, with diffuse emission having been subtracted, but accounting only for the obscured part of star formation \citep{lawton2010a,li2010a}. A dedicated study to disentangle the respective impact of the metallicity and the diffuse emission on the scaling factors at various scales would be required to fully understand this point.

\section{Obscured versus unobscured star formation\label{sec:att-frac}}

\subsection{FUV and H$\alpha$ attenuation in M33}

Because of the dust we only see a fraction of star formation in the UV or H$\alpha$. Following \cite{kennicutt2009a}, hybrid SFR estimators allow us to easily compute a proxy (noted $\mathcal{A}$) for the attenuation (noted A) of the UV and H$\alpha$ fluxes.
\begin{eqnarray}
 \mathcal{A}\mathrm{_{FUV}}&=&\mathrm{2.5\log\left[SFR(FUV+24~\mu m)/SFR(FUV)\right]},\\
 \mathcal{A}\mathrm{_{H\alpha}}&=&\mathrm{2.5\log\left[SFR(H\alpha+24~\mu m)/SFR(H\alpha)\right]}.\label{eqn:AHalpha}
\end{eqnarray}
We can also write this more directly in terms of luminosities:
\begin{eqnarray}
 \mathcal{A}\mathrm{_{FUV}}&=&\mathrm{2.5\log\left[1+k_{FUV-24}\times L(24~\mu m)/L(FUV)\right]},\\
 \mathcal{A}\mathrm{_{H\alpha}}&=&\mathrm{2.5\log\left[1+k_{H\alpha-24}\times L(24~\mu m)/L(H\alpha)\right]},
\end{eqnarray}
with $\mathrm{k_{band1-band2}}$ defined as in Sect.~\ref{ssec:SFR-estimators} and Table~\ref{tab:SFR}. These expressions can also be written equivalently in terms of surface brightnesses. Before going further, we should keep in mind that these estimators have been defined for star--forming regions and may not provide us with accurate estimates outside of their definition range.

\begin{figure*}[!htbp]
\includegraphics[width=\textwidth]{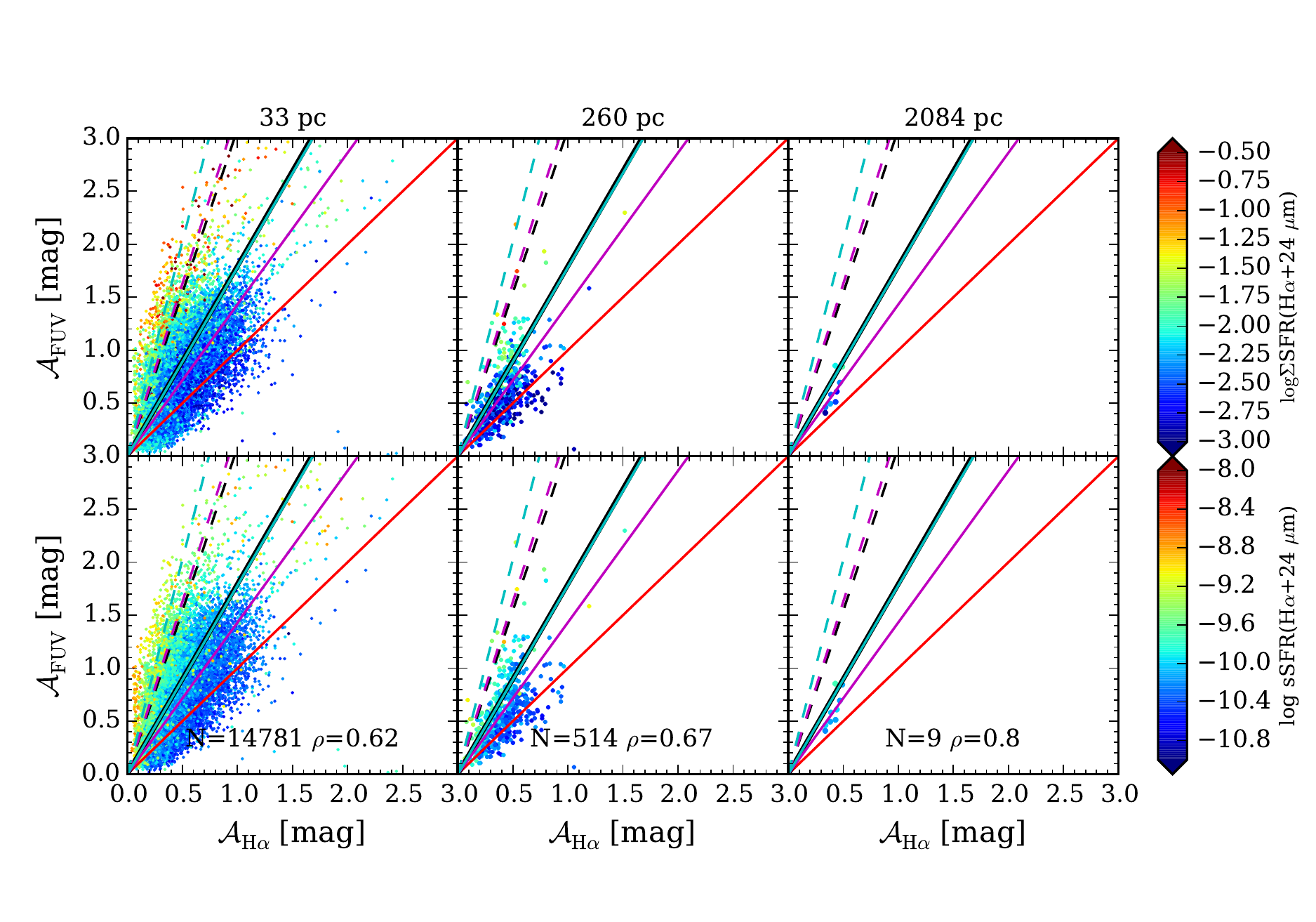}
\caption{Attenuation in the FUV band versus the attenuation in H$\alpha$. The colour of each point indicates $\Sigma$SFR (upper row) or the sSFR (lower row), following the bar to the right. In the bottom row, the number of regions N and the Spearman correlation coefficient $\rho$ are indicated. To compute the sSFR, the stellar mass in each region was computed from the 3.6~$\mu$m emission using the linear conversion factor of \cite{zhu2010a}. The red line shows the one--to--one relation. The black, magenta, and cyan lines represent the attenuation for a starburst, a Milky Way, and an LMC average curve with differential reddening ($f\equiv E(B-V)_{continuum}/E(B-V)_{gas}=0.44$, solid, with $E(B-V)_{continuum}$ being the reddening between the V and B bands of the stellar continuum and $E(B-V)_{gas}$ being that of the ionised gas) and without ($f=1$, dashed). For the starburst relation, we assumed that even though the stellar continuum follows the starburst curve, the gas still follows a Milky Way curve. Note that the black and cyan solid lines are nearly on top of one another. We find that at the finest resolution, there is a broad range in terms of differential reddening. Intense star--forming regions have little differential reddening, whereas diffuse regions on the contrary present a strong differential reddening. At coarser scales, the averaging between diffuse and star--forming regions yields a differential reddening that is similar to that of starburst galaxies. The overall shape of the attenuation law is however only weakly constrained and may vary across the galaxy.\label{fig:thumbnails-att-plots}}
\end{figure*}

Because the attenuation increases with decreasing wavelength, the attenuation in the FUV is higher than in the optical. For instance, if we consider the Milky Way extinction curve of \cite{cardelli1989a} with the \cite{odonnell1994a} update, for $\mathrm{A_V=1}$, $\mathrm{A_{H\alpha}\simeq0.8}$ and $\mathrm{A_{FUV}\simeq2.6}$. However, nebular emission is more closely linked to the most recent star formation episode, and therefore to dust, than the underlying stellar continuum. As a consequence, the H$\alpha$ line is actually more attenuated compared to the stellar continuum at the same wavelength than what we could expect from the extinction by a simple dust screen affecting both components the same way \citep{calzetti1994a,calzetti2000a,charlot2000a}. In reality this differential attenuation strongly depends on the geometry between the dust and the stars as well as on the star formation history. Given the broad range of physical conditions and scales in M33, we can expect the attenuation law between H$\alpha$ and the FUV band to vary strongly across the galaxy and across scales. Such variations would provide us with useful information on the effective attenuation curve between those two popular star formation tracers.
The relation between H$\alpha$ and FUV attenuations as a function of $\Sigma$SFR and the specific SFR (sSFR, the SFR per unit stellar mass) is shown in Fig.~\ref{fig:thumbnails-att-plots}.

We find that on average, the attenuation in M33 is relatively low for a spiral galaxy. There are peaks of attenuation reaching 2.5~mag in the FUV band at a resolution of 33~pc, but when we consider large sections of the galaxy at 2~kpc scales the typical attenuation is around 0.6~mag in the FUV band and 0.4~mag in H$\alpha$, making M33 mostly transparent in star formation tracing bands on large scales. While this is lower than the typical FUV attenuation in nearby spiral galaxies \citep{boquien2012a,boquien2013a}, it is consistent with previous findings in M33 \citep{tabatabaei2007b,verley2009a}. This difference compared to local spirals is probably due to the more metal--poor nature of M33.

Overall, we find that at the finest resolution, regions in M33 span a broad range in terms of absolute and relative attenuations in FUV and H$\alpha$. This does not appear to be due to random noise though as the locus of the regions appears structured according the the intensity of star formation. Regions with intense star formation as traced by the combination of H$\alpha$ with 24~$\mu$m, tend to have a high $\mathcal{A}\mathrm{_{FUV}}$ compared to $\mathcal{A}\mathrm{_{H\alpha}}$. This is especially visible at the finest spatial resolution. Intense star forming regions such as NGC~604 show a peak in $\mathcal{A}\mathrm{_{FUV}}$ whereas no particular increase is seen in $\mathcal{A}\mathrm{_{H\alpha}}$. If we select all pixels with $\mathrm{\Sigma SFR(H\alpha+24~\mu m)\geq0.1}$~M$_\sun$~yr$^{-1}$~kpc$^{-2}$ at 33~pc, we find $\left<\mathcal{A}\mathrm{_{FUV}}/\mathcal{A}\mathrm{_{H\alpha}}\right>=3.94\pm1.45$, versus $\left<\mathcal{A}\mathrm{_{FUV}}/\mathcal{A}\mathrm{_{H\alpha}}\right>=1.81\pm1.11$ for less active regions. As the resolution becomes coarser, excursions in attenuation become more moderate and the range covered in terms of FUV and H$\alpha$ attenuations becomes much smaller. At the coarsest resolution, $\mathcal{A}\mathrm{_{FUV}}$ and $\mathcal{A}\mathrm{_{H\alpha}}$ show little scatter  and they are consistent with a starburst or a Milky Way law with a differential reddening (see Sect.~\ref{ssec:variation-scale}) between the stellar continuum and the gas. What probably happens is that at coarser resolutions, intensely star-forming regions and quiescent regions merge together, decreasing the dynamic range in terms of attenuation properties. At the coarsest resolution, all regions have broadly similar properties, which is why they all have similar attenuation laws. We detail this aspect in Sect.~\ref{ssec:variation-scale}.

Finally, we should also mention the possibility that there is a change in the intrinsic extinction laws because of changes in the dust composition. Regions at low $\Sigma$SFR are located in the outskirts of the galaxy. However this is probably a minor effect. M33 has a very modest metallicity gradient of $-0.027\pm0.012$~dex/kpc \citep{rosolowsky2008a}. As we can see in Fig.~\ref{fig:thumbnails-att-plots}, a variation of the differential reddening has a much stronger effect than a change in the intrinsic extinction curve from the Milky Way to the LMC average.

\subsection{Variations of attenuation laws with scale\label{ssec:variation-scale}}

At first sight, these variations may seem at odds with the now well established picture of differential attenuation between the gas and the stars in galaxies \citep{calzetti1994a,calzetti2000a,charlot2000a}. However, this description was conceived in the particular context of starburst galaxies and may not apply directly to resolved and more quiescent galaxies. Let's first consider M33 at a resolution of 33~pc. As mentioned earlier, a low value for $\Sigma$SFR actually corresponds to diffuse emission with at most very little local star formation. In this environment, because gas is intimately linked with dust, the H$\alpha$ radiation always undergoes some attenuation. However, the stellar emission may be relatively attenuation--free as it is not particularly linked to dust, depending on the actual geometry. This would explain the relatively shallow effective FUV--H$\alpha$ attenuation curve that is normally seen in starburst galaxies. Now, if we consider star forming regions, the FUV--emitting stars will on average be younger and still closely linked to their birth cloud, hence undergoing a much higher attenuation than in diffuse regions. Because H$\alpha$ is always linked to dust, the increase of the attenuation is not as strong. If we now consider coarser resolutions, we increasingly mix diffuse and star--forming regions. At a local scale $\mathcal{A}\mathrm{_{FUV}}$ is on average much larger than $\mathcal{A}\mathrm{_{H\alpha}}$ in star--forming regions but more comparable in diffuse regions, as we have seen above. This means that at a global scale, the effective H$\alpha$--UV attenuation curve should be shallower than intrinsic extinction curves. This agrees with what we see at a scale of 2~kpc, $\left<\mathcal{A}\mathrm{_{FUV}}/\mathcal{A}\mathrm{_{H\alpha}}\right>=1.46\pm0.24$.

Using FUV to FIR broadband data on a sample of nearby, resolved galaxies at a typical scale of 1~kpc, \cite{boquien2012a} found hints of an evolution of the attenuation curve of the stellar continuum, with the age of star--forming regions, from a starburst--like curve in young regions to LMC--like curves in older regions. A consistent result was found on the scale of entire galaxies by \cite{kriek2013a}. They showed that $0.5<z<2.0$ galaxies with a high sSFR have a shallower attenuation curve. If we assume that a high $\Sigma$SFR is an indication of a young age, this would appear to be opposite of the trend we see in M33. However, a direct comparison is not straightforward because here we are comparing the nebular attenuation to the stellar continuum attenuation, and with measurements at only two wavelengths. In other words, we are looking at the difference between the gas and the stellar attenuation curves, measuring each at a single wavelength only.

A major and poorly constrained factor that is important for this comparison is the differential reddening we mentioned earlier, which we can write as: $f=\mathrm{E(B-V)_{ continuum}/E(B-V)_{gas}}$. This can also be expressed in terms of attenuations. Considering that $\mathrm{E(B-V)=A_V/R_V}$, $f=\mathrm{A_{V,continuum}/A_{V,gas}\times R_{V,gas}/R_{V,continuum}}$. As we have stated earlier, in diffuse regions FUV--emitting stars are probably more weakly linked to the dust than the ionised gas. As such, in diffuse regions $f$ may be much smaller than what it is in pure star--forming regions where it should be closer to $f=1$. This means that in diffuse regions the attenuation of the stellar continuum would be much smaller than the attenuation of the nebular emission at a given wavelength. Based on a sample of galaxies observed by the SDSS, \cite{wild2011a} found that the optical depth of nebular emission compared to that of the continuum is significantly higher for galaxies at low sSFR. They attributed this to a variation of the relative weight of diffuse and star--forming regions. This means that $f$ is smaller in these more quiescent galaxies. Similar results have been obtained by \cite{price2014a} based on the 3D--HST survey and by \cite{kashino2013a} using ground--based spectra of galaxies at $z=1.6$. To verify these results in M33, in Fig.~\ref{fig:thumbnails-att-plots} we have also colour coded the relation between $\mathcal{A}\mathrm{_{FUV}}$ and $\mathcal{A}\mathrm{_{H\alpha}}$ as a function of the sSFR. We find a result consistent with that of the aforementioned works. Regions with a high sSFR have a high value of $f$ whereas regions with a low sSFR have a low value of $f$. This way, considering a variation of $f$, it is possible that the effective FUV--H$\alpha$ attenuation curve would show a different evolution compared to the attenuation curve of the stellar continuum emission. Our results suggest both a variation of $f$ across the galaxy at a given scale from diffuse regions to star--forming regions, and a variation depending on the scale due to averaging of star--forming and diffuse regions that have different values of $f$. At the finest resolution, a range of $f$ is required to explain the observations across the galaxy. However, as we go towards coarser resolutions, the observations can be explained with $f=0.44$.

\subsection{Limits on the determination of the attenuation}

This discussion relies on the assumption that no systematic bias is introduced due to the way we compute the attenuation and $\Sigma$SFR. If we consider $\Sigma$SFR from FUV and 24~$\mu$m rather than from H$\alpha$ and 24~$\mu$m, the trends are not as clear. There is a fraction of pixels at 33~pc with very low FUV attenuation ($\mathrm{0\lesssim \mathcal{A}\mathrm{_{FUV}}\lesssim0.1}$) and moderately high $\Sigma$SFR. This probably corresponds to regions with a low level of 24~$\mu$m and H$\alpha$ emission but with strong FUV. That could be the case for instance in a region where recently formed clusters have blown away much of the dust and the gas of their parent clouds. Such regions were found by \cite{relano2013a}, especially in the outskirts of M33.

A specific bias may affect some diffuse regions. The most extreme have $\mathcal{A}\mathrm{_{FUV}}<\mathcal{A}\mathrm{_{H\alpha}}$, which would require a particularly strong differential attenuation. A close inspection reveals that these regions are also relatively fainter in H$\alpha$. The relatively higher uncertainties would then propagate into the attenuation estimates yielding spuriously low $\mathcal{A}\mathrm{_{H\alpha}}$. In practice, they could also be affected by very strong age and radiation transfer effects such as the escape of ionising photons, which would reduce the local H$\alpha$ luminosity, independent of the actual attenuation underwent by H$\alpha$ photons. In that case, with little H$\alpha$ compared to the FUV and for the same amount of 24~$\mu$m, the selected estimators will then naturally overestimate $\mathrm{A_{H\alpha}}$. These regions would in reality not present a differential attenuation as extreme as could be inferred from our estimates. To ensure that these uncertainties on diffuse regions do not affect our results, we have selected only regions with $\mathrm{\Sigma SFR(H\alpha+24~\mu m)>10^{-2}}$~M$_\odot$~yr$^{-1}$~kpc$^{-2}$, hence removing purely diffuse regions. {We find that we still see the clear gradients described in Fig.~\ref{fig:thumbnails-att-plots}. This means that if in the most extreme regions, the differential attenuation is likely to be overestimated, there is still a clear variation of the differential attenuation depending on the sSFR. 

The issues we have presented show the sensitivity of such an analysis on the selected SFR estimators and the great caution that must be used when interpreting such results. A promising way to reduce such potential problems would be to compute the attenuation with a full SED modelling for the stellar continuum and from the Balmer decrement for the nebular emission. The increasing availability of spectral maps using integral field spectrographs (IFS), and large multi--wavelengths surveys now makes this possible for nearby galaxies \citep[e.g.,][]{sanchez2012a,blanc2013a}. Recently, \cite{kreckel2013a} have used such IFS data on a sample of 8 nearby galaxies, deriving the nebular attenuation from the Balmer decrement and the stellar attenuation from the shape of the continuum between 500~nm and 700~nm. Interestingly, opposite to our results and that of \cite{wild2011a}, they find that in diffuse regions the attenuation of the stars increases compared to that of the gas. In the most extreme cases, in the V band the stellar attenuation is 10 times higher than that of the gas. Conversely, for regions with $\mathrm{\Sigma SFR>10^{-1}}$~M$_\odot$~yr$^{-1}$~kpc$^{-2}$, they converge to $f=0.47$, close to what we find at the coarsest resolution. The discrepancy at low $\mathrm{\Sigma SFR}$ may be due to systematics in the way the attenuation is computed for the diffuse medium, both for the stars and the gas. For similar sized regions at high $\mathrm{\Sigma SFR}$, the discrepancy is probably due to the fact we measure the continuum attenuation in the FUV whereas \cite{kreckel2013a} measure it in the optical. In their case, even in star forming regions the continuum emission is generally dominated by older stellar populations, which is not necessarily the case in the FUV, inducing a different $f$. This effect is probably prevalent mainly at the smallest scales

To summarise, great care must be used when correcting star--formation tracing bands for the attenuation. We have shown that there are clear variations of the effective FUV-H$\alpha$ attenuation curves that depend on the sSFR and $\Sigma$SFR, with regions at higher SFR having steeper attenuation curves. This is due to a strong variation of the differential reddening between the stars and the gas. Intense star--forming regions have little differential reddening ($f\simeq1$), contrary to more quiescent regions. Finally, there is also a strong variation with the resolution, due to averaging regions with different physical properties. At the coarsest resolution, the effective attenuation curve is compatible a differential reddening of $f=0.44$, which is the value for the starburst curve for instance. However, it is not possible to discriminate between different laws at fixed differential reddening.

\section{Discussion\label{sec:discussion}}

We have found that there is an important variation in the differential attenuation in M33 with both the spatial scale and the sSFR. At the same time, it appears that resolution effects become small beyond a scale of 1~kpc. In light of these results, we now have a better insight into the relation between UV--emitting stars and dust in galaxies (Sect.~\ref{ssec:geometry}). They also allow us to understand how the measure of the SFR will be affected by high resolution observations with upcoming instruments (Sect.~\ref{ssec:impact}).

\subsection{Constraints on the relative geometry of stars and dust in star--forming galaxies\label{ssec:geometry}}

The actual geometry between the stars and the dust in galaxies is undoubtedly complex, but a simple generalised model has emerged for starburst and more quiescent star--forming galaxies \citep{calzetti1994a,wild2011a,price2014a}. These descriptions generally rely on a two--component model framework \citep[e.g.,][]{charlot2000a}: dense star--forming regions and a lower density diffuse medium. We will not come back to the general descriptions of galaxies that have been discussed in detail in the literature \citep[e.g.,][]{wild2011a}. Our multi--scale analysis however sheds light on the distribution between FUV--emitting stars and the dust at a local scale.

In diffuse regions we have found that the differential reddening is large. This shows that the FUV--emitting stars are relatively unassociated with dust. This requires that these stars have escaped their birth cocoon or that stellar feedback has induced a physical displacement between the young stars one hand and the gas and the dust on the other hand. Conversely, the nebular emission is more strongly attenuated. This means that the ionised gas is more associated with dust than the stars. Several mechanisms can be invoked. First, this emission may originate from gas ionised by nearby massive stars or created by ionising radiation that has escaped from more distant star--forming regions. \cite{hoopes2000a} find that massive stars in the field can account for 40\% of the ionisation of the diffuse ionised gas in M33. An alternative is that it comes from H$\alpha$ photons that have travelled a long distance in the plane of the disk before being scattered in the direction of the line of sight. The latter possibility is less likely as it would locally boost the H$\alpha$ luminosity relative to the 24~$\mu$m one, thereby reducing the attenuation inferred from Eq.~\ref{eqn:AHalpha}.

Conversely, in star--forming regions there is very little differential reddening. This suggests that the UV--emitting stars, the dust, and the gas are well mixed and follow similar distributions. The actual geometry drives the transformation of the extinction curve, which describes the case when there is a simple dust screen in front of the sources, into an attenuation curve. Constraining the geometry would require additional data to attempt to break the various degeneracies affecting the determination of the attenuation curve. This is a notoriously difficult task, especially since the structure of the ISM is much more complex than the simple assumptions that are usually made.

The progressive convergence towards the canonical differential reddening of $f=0.44$ at larger scales shows the impact of the distribution of gas and stars at local scales on the galaxy seen at coarser scales. But this also shows the danger of assuming similar geometries and attenuation curves across all scales and all regions in resolved galaxies. Assuming a differential reddening different from what it is in reality can lead to errors of a factor several on the determination of the attenuation, and therefore on the determination of the SFR. In other words: there is no one attenuation law that is valid under all circumstances. However, considering regions of at least 1~kpc across strongly limits resolution effects to compute the SFR. This is probably due to the broad mixing between star--forming and diffuse regions. We will explore in the next section when this scale dependence is most likely to have an impact in the era of high resolution observations.

We present a simplified graphical description of the relative geometry of stars, gas, and dust in diffuse and in star--forming regions in Fig.~\ref{fig:geometry}.
\begin{figure*}[!htbp]
\includegraphics[width=\textwidth]{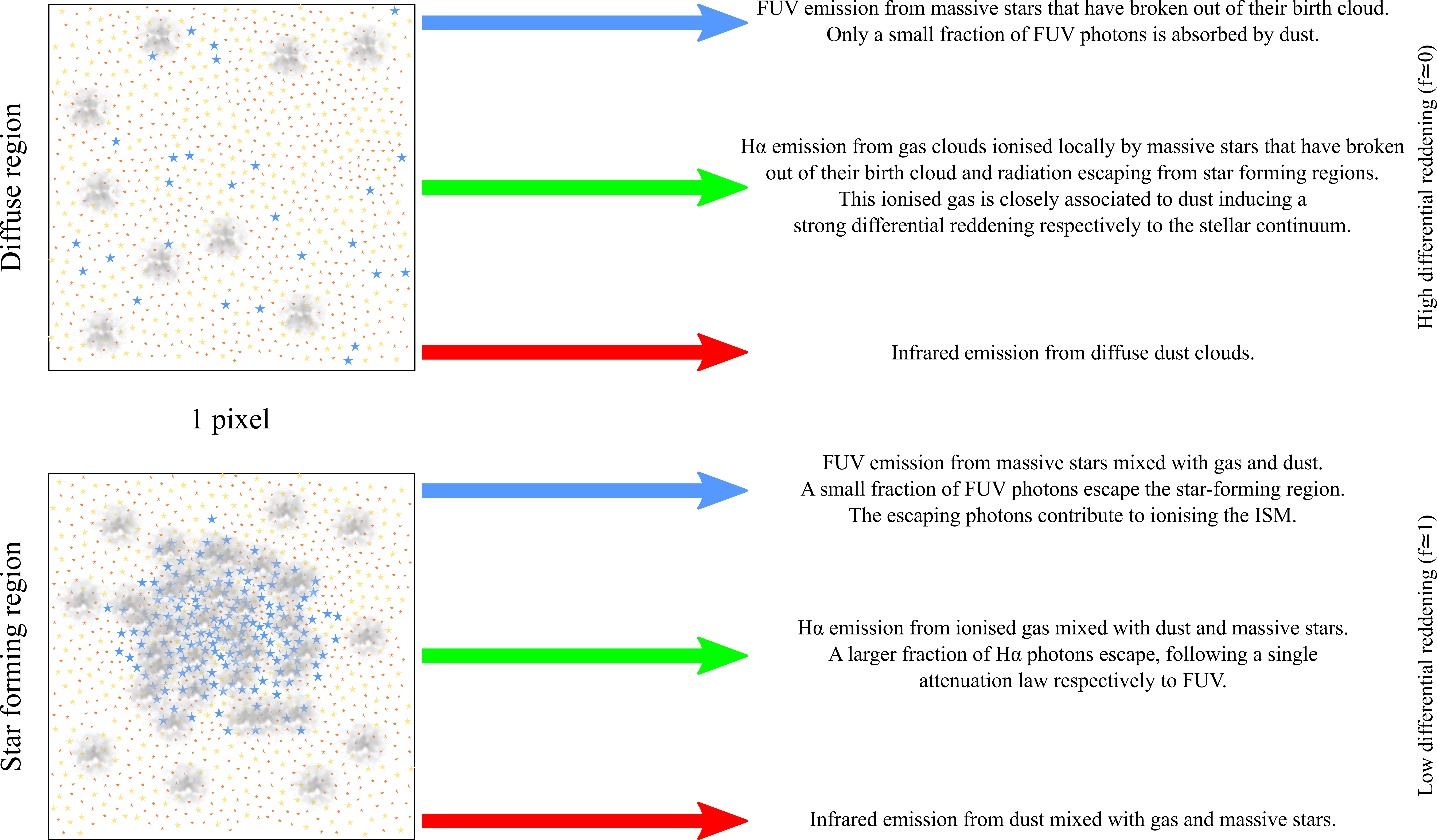}
\caption{Simplified description of the relative distribution of stars, gas, and dust at a local scale in diffuse regions (top) and in star--forming regions (bottom). The FUV emitting stars are shown in blue whereas older stellar populations are in yellow or orange. The clouds of gas and dust are symbolised with grey patches.\label{fig:geometry}}
\end{figure*}
It is conceptually similar to Fig.~8 in \cite{calzetti2001a} but at the same time it shows the fundamental difference between normal star--forming galaxies and starburst galaxies.

\subsection{Measuring high redshift star formation in the era of high resolution ALMA and the JWST observations\label{ssec:impact}}

As we have shown across this article, the determination of the SFR or the attenuation is not only luminosity dependent, but it is also scale dependent. With the recent commissioning of ALMA and the launch of the JWST by the end of the decade, it will finally be possible to carry out highly resolved observations of star formation not only in the nearby universe but also well beyond. With such opportunities also come the complexities inherent to high resolution studies. To examine in which cases the interpretation of the observations may be affected by the resolution, we have plotted in Fig.~\ref{fig:resolution} the physical scale that can be reached in the UV and in the IR with the JWST and ALMA.
\begin{figure}[!htbp]
\includegraphics[width=\columnwidth]{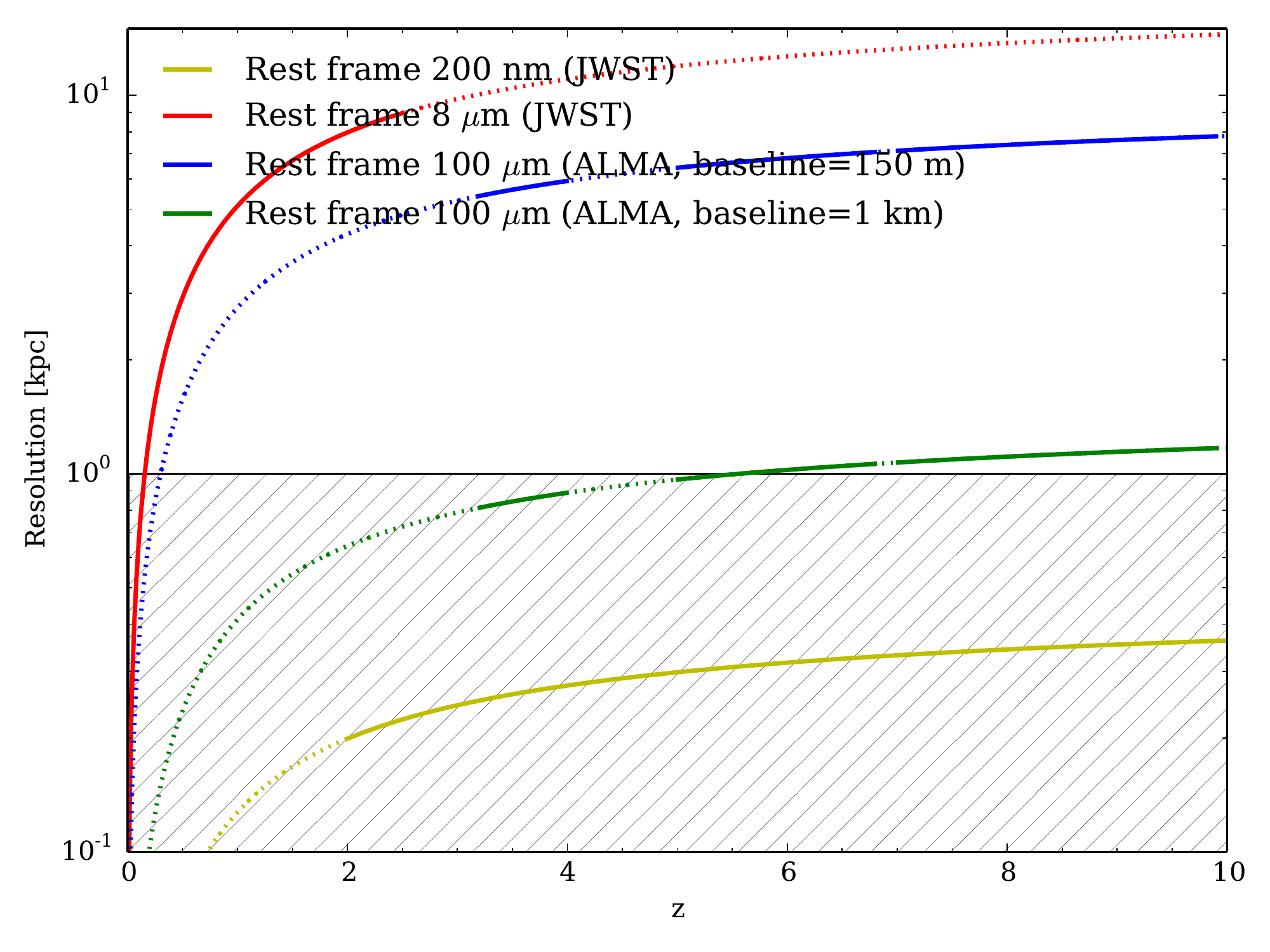}
\caption{Spatial resolution versus $z$ for a rest frame wavelength of 200~nm (yellow) and 8~$\mu$m (red) with the JWST, and at 100~$\mu$m with ALMA for baselines of 150~m (blue) and 1~km (green). The \cite{planck2013_16} cosmological parameters are adopted. The solid lines show at which redshifts the observations can be carried out. With the JWST Observations below 200~nm are not possible under $z=2$, while observations beyond $z=2.5$ are not possible at 8~$\mu$m. Conversely, the rest frame emission at 100~$\mu$m cannot be observed with ALMA below $z=3.2$. There are several gaps at longer wavelengths corresponding to the gaps between different ALMA bands. These bands correspond to the ones available for cycle 2. Band 10 will strongly improve the capabilities of ALMA to map the main infrared star formation bands at moderate redshifts while band 11 will allow us to probe the peak of dust emission down to $z=1$. Finally, the hatched area corresponds to a resolution smaller than 1~kpc, where there may be a strong impact on the measure of star formation.\label{fig:resolution}}
\end{figure}

Unsurprisingly, the highest resolution will be achieved in the rest--frame 200~nm, which will allow us to distinguish 500~pc details all the way to $z=10$. Unless degraded to lower resolution, these images may prove problematic to derive reliably the local physical parameters. Conversely, the resolution of rest--frame 8~$\mu$m rapidly degrades with increasing redshift, reaching 9~kpc at the maximum redshift of $z=2.5$. It would still be considerably useful to carry out resolved studies of low--redshift galaxies, a resolution of 1~kpc is already achieved at $z=0.15$. 

To probe the peak of dust emission at 100~$\mu$m with ALMA, a baseline of 1~km appears nearly perfect, with the resolution only slightly varying around 1~kpc from $z=3.2$ to $z=9.9$. A baseline of only 150~m would only provide us with a much coarser resolution of 4 to 5~kpc. An additional complexity not taken into account here would be the loss of $uv$ coverage from the lack of short baselines, which would be especially problematic at low redshift. This would require complementary observations with the Atacama Compact Array. The addition in the future of bands 10 and perhaps 11 will allow the use of shorter baselines while extending the window to lower redshift galaxies. Band 11 would be able to detect 100~$\mu$m emission down to $z=1$.

Overall, the synergy between ALMA and the JWST is excellent to probe star formation at a well resolved scale while also gaining valuable insight into triggering or feedback. The combination of these instruments will extend to much higher redshifts the spatially resolved multi--wavelength studies that can only be done on nearby galaxies currently.

\section{Conclusion\label{sec:conclusion}}

In order to understand how SFR measurements of galaxies depend on the physical scale, we have carried out an analysis of the emission of the local group galaxy M33 from 33~pc to 2084~pc. We have found that:
\begin{enumerate}
 \item Monochromatic SFR estimators can be strongly discrepant compared to a reference H$\alpha$+24~$\mu$m estimator. These discrepancies depend on the scale of the study and on $\Sigma$SFR. They may be due to be combined effects of the age, the geometry, the transparency of the ISM, and the importance of diffuse emission.
 \item The scaling factors from individual infrared bands to $\Sigma$SFR show an important evolution with the physical size, up to a factor 2. Star--forming and diffuse regions show a different evolution with the spatial scale. There is however a convergence of scaling factors at large scales.
 \item More generally, such variations with the physical scale and the discrepancies of the scaling relations compared to those obtained from different samples show that it is especially dangerous to apply SFR estimators beyond their validity range in terms of surface brightness, physical scale, and metallicity. This issue is especially important when applying SFR estimators on higher redshift galaxies as their physical properties may be more poorly known. This is why we make no attempt at deriving a multi--scale SFR estimator as it would be strongly tied with M33. That being said, carrying out studies at a scale coarser than 1~kpc strongly limits resolution effects. Such resolutions will be routinely achieved at high redshift with ALMA and the JWST.
 \item Finally, there is a clear change in the differential reddening between the nebular emission and the stellar continuum depending both on the physical scale and on $\Sigma$SFR or the sSFR. Star--forming regions have nearly no differential reddening whereas diffuse regions have a strong differential reddening. Such a change in the reddening is especially visible at the finest spatial resolution. At coarser resolutions, the differential reddening converges to values compatible with the canonical 0.44 value derived for starburst galaxies by \cite{calzetti2000a}. These results allow us to obtain new insights into the relative geometry between the stars and the dust at a local scale in galaxies, from diffuse regions to star--forming regions.
\end{enumerate}

The maps presented in this article are available in the form of a FITS file at the following address: \url{http://www.ast.cam.ac.uk/~mboquien/m33/}.

\begin{acknowledgements}
MB thanks Robert Kennicutt, Kathryn Kreckel, Yiming Li, and Vivienne Wild for useful discussions that have helped improve the paper.

This research has made use of the NASA/IPAC Extragalactic Database (NED) which is operated by the Jet Propulsion Laboratory, California Institute of Technology, under contract with the National Aeronautics and Space Administration. 

This research has made use of the NASA/IPAC Infrared Science Archive, which is operated by the Jet Propulsion Laboratory, California Institute of Technology, under contract with the National Aeronautics and Space Administration.

This research made use of Astropy, a community-developed core Python package for Astronomy \citep{astropy2013a}.

This research made use of APLpy, an open-source plotting package for Python hosted at \url{http://aplpy.github.com}.
\end{acknowledgements}

\bibliographystyle{aa}
\bibliography{article}
\end{document}